\documentclass[aps,prd,twocolumn,showpacs]{revtex4-1}

\usepackage{wallpaper}
\usepackage[colorlinks,plainpages=false]{hyperref}
\usepackage{amsmath}
\usepackage{multirow}

\begin{document}
\hyphenpenalty=6000
\tolerance=1000

\hyphenation{Eq}

\title{Nuclear pasta structures and symmetry energy}
\author{Cheng-Jun~Xia$^{1,2}$}
\email{cjxia@nit.zju.edu.cn}
\author{Toshiki Maruyama$^{2}$}
\email{maruyama.toshiki@jaea.go.jp}
\author{Nobutoshi Yasutake$^{3, 2}$}
\email{nobutoshi.yasutake@it-chiba.ac.jp}
\author{Toshitaka Tatsumi$^{4}$}
\email{tatsumitoshitaka@gmail.com}
\author{Ying-Xun Zhang$^{5, 6}$}
\email{zhyx@ciae.ac.cn}

\affiliation{$^{1}${School of Information Science and Engineering, NingboTech University, Ningbo 315100, China}
\\$^{2}${Advanced Science Research Center, Japan Atomic Energy Agency, Shirakata 2-4, Tokai, Ibaraki 319-1195, Japan}
\\$^{3}${Department of Physics, Chiba Institute of Technology (CIT), 2-1-1 Shibazono, Narashino, Chiba, 275-0023, Japan}
\\$^{4}${Institute of Education, Osaka Sangyo University, 3-1-1 Nakagaito, Daito, Osaka 574-8530, Japan}
\\$^{5}${China Institute of Atomic Energy, Beijing 102413, People's Republic of China}
\\$^{6}${Guangxi Key Laboratory Breeding Base of Nuclear Physics and Technology, Guilin 541004, China}}

\date{\today}

\begin{abstract}
In the framework of the relativistic mean field model with Thomas-Fermi approximation, we study the structures of low density nuclear matter in a three-dimensional geometry with reflection symmetry. The numerical accuracy and efficiency are improved by expanding the mean fields according to fast cosine transformation and considering only one octant of the unit cell. The effect of finite cell size is treated carefully by searching for the optimum cell size. Typical pasta structures (droplet, rod, slab, tube, and bubble) arranged in various crystalline configurations are obtained for both fixed proton fractions and $\beta$-equilibration. It is found that the properties of droplets/bubbles are similar in body-centered cubic (BCC) and face-centered cubic (FCC) lattices, where the FCC lattice generally becomes more stable than BCC lattice as density increases. For the rod/tube phases, the honeycomb lattice is always more stable than the simple one. By introducing an $\omega$-$\rho$ cross coupling term, we further examine the pasta structures with a smaller slope of symmetry energy $L = 41.34$ MeV, which predicts larger onset densities for core-crust transition and non-spherical nuclei. Such a variation due to the reduction of $L$ is expected to have impacts on various properties in neutron stars, supernova dynamics, and binary neutron star mergers.
\end{abstract}

\maketitle

\section{\label{sec:intro}Introduction}
Due to the first-order liquid-gas phase transition of nuclear matter, a mixed phase is expected at subsaturation densities ($n_\mathrm{b}\lesssim 0.08\ \mathrm{fm}^{-3}$) and small temperatures ($T\lesssim 10$ MeV). Such a liquid-gas mixed phase will exhibit various nonuniform structures that are usually referred to as nuclear pasta~\cite{Baym1971_ApJ170-299, Negele1973_NPA207-298, Ravenhall1983_PRL50-2066, Hashimoto1984_PTP71-320, Williams1985_NPA435-844}, which exist typically in the inner crust region of neutron stars and the core region of supernovae at the stage of gravitational collapse. A detailed investigation on the possible structures and properties of nuclear pasta is essential to understand the rotation and thermal evolution of neutron stars~\cite{Lorenz1993_PRL70-379, Mochizuki1995_ApJ440-263, Levin2001_MNRAS324-917, Horowitz2004_PRC69-045804, Gusakov2004_AA421-1143, Gearheart2011_MNRAS418-2343, Pons2013_NP9-431, Rueda2014_PRC89-035804, Carvalho2014_PRC90-055804, Watanabe2017_PRL119-062701, Sotani2019_MNRAS489-3022, Schuetrumpf2020_PRC101-055804, Pethick2020_PRC101-055802}, supernova dynamics~\cite{Bethe1990_RMP62-801, Watanabe2005_PRL94-031101, Alloy2011_PRC83-035803, Roggero2018_PRC97-045804, Janka2012_ARNPS62-407}, and binary neutron star mergers~\cite{Gamba2019_CQG37-025008, Biswas2019_PRD100-044056, Baiotti2019_PPNP109-103714, LI2020, Gittins2020_PRD101-103025}.

In the past few decades, significant efforts were devoted to determine the structures of the nuclear pasta. For example, employing spherical and cylindrical approximations of the Wigner-Seitz (WS) cell~\cite{Pethick1998_PLB427-7, Oyamatsu1993_NPA561-431, Maruyama2005_PRC72-015802, Togashi2017_NPA961-78, Shen2011_ApJ197-20}, it was found that there exist five types of geometrical structures, i.e, droplets/bubbles, rods/tubes, and slabs for three, two, and one dimensions. Owing to the geometrical symmetry, the numerical calculation is essentially one-dimensional. However, such approximations neglect the interactions among other cells and thus have no dependence on the lattice type~\cite{Oyamatsu1984_PTP72-373}.
Meanwhile, further investigations have revealed much more complicated structures~\cite{Magierski2002_PRC65-045804, Newton2009_PRC79-055801, Fattoyev2017_PRC95-055804}, such as the gyroid and double-diamond morphologies~\cite{Nakazato2009_PRL103-132501, Schuetrumpf2015_PRC91-025801}, P-surface configurations~\cite{Schuetrumpf2013_PRC87-055805, Schuetrumpf2019_PRC100-045806}, nuclear waffles~\cite{Schneider2014_PRC90-055805, Sagert2016_PRC93-055801}, Parking-garage structures~\cite{Berry2016_PRC94-055801}, deformations in droplets~\cite{Kashiwaba2020_PRC101-045804}, as well as the intermediate structures of droplet and rod, slab and tube~\cite{Watanabe2003_PRC68-035806, Okamoto2012_PLB713-284}, which can not be described in the spherical or cylindrical approximations of the WS cell. Under such circumstances, in this work we investigate the nuclear pasta in a three-dimensional geometry with reflection symmetry, where the Thomas-Fermi approximation is adopted~\cite{Okamoto2012_PLB713-284, Okamoto2013_PRC88-025801}. The lattice structure, interaction among different unit cells, and charge screening effect can then be considered self-consistently. As was done in Refs.~\cite{Maruyama2005_PRC72-015802, Avancini2008_PRC78-015802, Avancini2009_PRC79-035804, Gupta2013_PRC87-028801}, the local properties of nuclear matter are obtained in the framework of the relativistic mean field (RMF) model~\cite{Meng2016_RDFNS}.

The nuclear matter properties are well constrained around the saturation density ($n_0\approx 0.16\ \mathrm{fm}^{-3}$) according to various terrestrial experiments and nuclear theories~\cite{Dutra2014_PRC90-055203}, which gives the binding energy $B\approx -16$ MeV, the incompressibility $K = 240 \pm 20$ MeV~\cite{Shlomo2006_EPJA30-23}, the symmetry energy $S = 31.7 \pm 3.2$ MeV and its slope $L = 58.7 \pm 28.1$ MeV~\cite{Li2013_PLB727-276, Oertel2017_RMP89-015007}. Note that the uncertainty of $L$ is larger than other quantities, which is expected to be reduced with the measurements of neutron skin thickness $\Delta R_{np}$ in the pioneering Lead Radius Experiment (PREX) II~\cite{PREX2021} and the upcoming Mainz Radius Experiment (MREX). At this moment, the neutron skin thickness of $^{208}$Pb measured in PREX-I is $\Delta R_{np}=0.33^{+0.16}_{-0.18}$ fm~\cite{PREX2012_PRL108-112502}, while a recent measurement with PREX-II suggests $\Delta R_{np}=0.283\pm 0.071$ fm~\cite{PREX2021}. The uncertainty can be reduced if other constraints are included~\cite{Dong2015_PRC91-034315, Roca-Maza2015_PRC92-064304, Fattoyev2018_PRL120-172702, Horowitz2019_AP411-167992, Essick2021}. It is worth mentioning that the symmetry energy at baryon number density $n_\mathrm{b}=0.1\ \mathrm{fm}^{-3}$ is well constrained ($25.5 \pm 1.0$ MeV) by reproducing finite nuclei properties~\cite{Centelles2009_PRL102-122502, Brown2013_PRL111-232502}, while its slope was shown to be deeply connected with $\Delta R_{np}$~\cite{Zhang2013_PLB726-234}.

Meanwhile, as we are entering the multi-messenger era, significant progresses were made on measuring neutron star properties~\cite{Lattimer2012_ARNPS62-485, Ozel2016_ARAA54-401, Baiotti2019_PPNP109-103714, Weih2019_ApJ881-73}. The precise mass measurements of the two-solar-mass pulsars~\cite{Demorest2010_Nature467-1081, Antoniadis2013_Science340-1233232, Fonseca2016_ApJ832-167, Cromartie2020_NA4-72} and the possible existence of more massive pulsars~\cite{Linares2018_ApJ859-54, Rezzolla2018_ApJ852-L25, Ruiz2018_PRD97-021501, Shibata2019_PRD100-023015, LVC2020_ApJ896-L44} have put strong constraints on the properties of dense stellar matter. With pulse-profile modeling~\cite{Watts2019_SCPMA62-29503}, recently the mass and radius of PSR J0030+0451 are accurately measured~\cite{Riley2019_ApJ887-L21, Miller2019_ApJ887-L24}. Nevertheless, the most stringent constraints on radius are obtained from the binary neutron star merger event {GRB} 170817A-{GW}170817-{AT} 2017gfo ($11.9^{+1.4}_{-1.4}$ km)~\cite{LVC2018_PRL121-161101}, corresponding to the measured dimensionless combined tidal deformability  $302\leq \tilde{\Lambda} \leq 720$~\cite{LVC2017_PRL119-161101, LVC2019_PRX9-011001, Coughlin2019_MNRAS489-L91, Carney2018_PRD98-063004, De2018_PRL121-091102, Chatziioannou2018_PRD97-104036}. The uncertainties of nuclear matter properties can be further reduced adopting those constraints~\cite{Tsang2019_PLB795-533}. In fact, it was shown that the radius and tidal deformability of neutron stars are closely related to $L$~\cite{Zhu2018_ApJ862-98, Tsang2019_PLB795-533, Dexheimer2019_JPG46-034002, Zhang2019_EPJA55-39, Zhang2020_PRC101-034303, Li2020_PRC102-045807}. By combining all these constraints and the heavy ion collision data, a recent estimation using the effective Skyrme energy density functional suggests $K = 250.23 \pm 20.16$ MeV, $S = 31.35 \pm 2.08$ MeV and $L = 59.57  \pm 10.06$ MeV~\cite{Zhang2020_PRC101-034303}.

In light of the updated constraints on nuclear matter properties and particularly the slope of symmetry energy, we reanalyze the results obtained in previous study~\cite{Okamoto2012_PLB713-284, Okamoto2013_PRC88-025801} and introduce an $\omega$-$\rho$ cross coupling term. The slope of symmetry energy is then reduced from $L = 89.39$ MeV~\cite{Maruyama2005_PRC72-015802} to $L = 41.34$ MeV, which better reproduces the neutron star tidal deformability. Previous studies adopting the spherical and cylindrical approximations of the WS cell suggest that the charge number of nuclei, the core-crust transition density, and the onset density of non-spherical nuclei decrease with $L$~\cite{Oyamatsu2007_PRC75-015801, Grill2012_PRC85-055808, Bao2015_PRC91-015807, Shen2020_ApJ891-148}. For the core-crust transition density $n_\mathrm{t}$, similar $n_\mathrm{t}$-$L$ relation was found using both the dynamical and thermodynamical methods~\cite{Xu2009_ApJ697-1549}, while recently it was suggested that higher order terms of symmetry energy could also play important roles~\cite{Pais2012_PRL109-151101, Li2020_PRC102-045807}.

In this work we thus examine the impact of varying $L$ on nuclear pasta, where various crystal structures are considered. In order to determine the true ground state with higher accuracy, we expand the mean fields according to fast cosine transformation. The computation time is further reduced by considering only one octant of the unit cell~\cite{Newton2009_PRC79-055801}. The effect of finite cell size~\cite{GimenezMolinelli2014_NPA923-31, Newton2009_PRC79-055801} is then accounted for by searching for the optimum cell size. The paper is organized as follows. In Sec.~\ref{sec:the}, we present our theoretical framework of RMF model. In Sec.~\ref{sec:num} the $\omega$-$\rho$ and $N$-$\rho$ coupling constants are adjusted according to various constraints, while the numerical details on obtaining the nuclear pasta are discussed. The obtained results on the structure and properties of nuclear pasta are presented in Sec.~\ref{sec:pasta}. Our conclusion is given in Sec.~\ref{sec:con}.

\section{\label{sec:the} Theoretical framework}

The Lagrangian density of the RMF model~\cite{Meng2016_RDFNS} reads
\begin{eqnarray}
\mathcal{L}
 &=& \sum_{i} \bar{\psi}_i \left[ i \gamma^\mu \partial_\mu - m_i - g_{i\sigma} \sigma \right. \nonumber \\
 &&\mbox{}\left. - \gamma^\mu \left( g_{i\omega}  \omega_\mu + g_{i\rho} \boldsymbol{\tau}_i \cdot \boldsymbol{\rho}_\mu
              + q_i A_\mu\right)    \right] \psi_i  \nonumber \\
 &&\mbox{} + \frac{1}{2}\partial_\mu \sigma \partial^\mu \sigma - \frac{1}{2}m_\sigma^2 \sigma^2
     - U(\sigma) - \frac{1}{4} \omega_{\mu\nu}\omega^{\mu\nu} \nonumber \\
 &&\mbox{}+ \frac{1}{2}m_\omega^2 \omega_\mu\omega^\mu
     - \frac{1}{4} \boldsymbol{\rho}_{\mu\nu}\cdot\boldsymbol{\rho}^{\mu\nu}
     + \frac{1}{2}m_\rho^2 \boldsymbol{\rho}_\mu\cdot\boldsymbol{\rho}^\mu \nonumber \\
 &&\mbox{} - \frac{1}{4} A_{\mu\nu}A^{\mu\nu}
     + \Lambda_\mathrm{v}g_\omega^2 g_\rho^2 (\omega_\mu\omega^\mu) (\boldsymbol{\rho}_\mu\cdot\boldsymbol{\rho}^\mu).
\label{eq:Lagrange}
\end{eqnarray}
Here the Dirac spinor $\psi_i$ represents a fermion ($n$, $p$, $e$) with mass $m_i$, isospin $\boldsymbol{\tau}_i$, and charge $q_i$. Three types of mesons ($\sigma$, $\omega^\mu$, and $\boldsymbol{\rho}^\mu$) are included to account for the strong interactions among nucleons, where the nucleon-meson coupling constants are taken as $g_{n\sigma} = g_{p\sigma} = g_{\sigma}$, $g_{n\omega} = g_{p\omega} = g_{\omega}$, and $g_{n\rho} = g_{p\rho} = g_{\rho}$. We take $g_{e\sigma} = g_{e\omega} = g_{e\rho} = 0$ since electrons have nothing to do with strong interaction. To account for the density dependence of effective interaction strengths, we adopt the nonlinear self-couplings of $\sigma$, i.e.,
\begin{equation}
  U(\sigma) = b m_N (g_{\sigma} \sigma)^3/3 + c (g_{\sigma} \sigma)^4/4.
\end{equation}
Meanwhile, an $\omega$-$\rho$ cross coupling term $\Lambda_\mathrm{v}g_\omega^2 g_\rho^2 (\omega_\mu\omega^\mu) (\boldsymbol{\rho}_\mu\cdot\boldsymbol{\rho}^\mu)$ is included here to give better constraints on the density dependence of symmetry energy~\cite{Shen2020_ApJ891-148}. If we fix the symmetry energy at baryon number density $n_\mathrm{b}=0.11\ \mathrm{fm}^{-3}$ by readjusting $g_\rho$, it was shown that the slope of symmetry energy $L$ decreases with $\Lambda_\mathrm{v}$~\cite{Bao2015_PRC91-015807}. In principle, one could also introduce other cross coupling terms such as $\sigma$-$\rho$ and $\sigma$-$\omega$ terms in the Lagrangian density~\cite{Dutra2014_PRC90-055203}. Alternatively, adopting the Typel-Wolter ansatz~\cite{Typel1999_NPA656-331}, an explicit density dependent $N$-$\rho$ coupling constant $g_\rho$ can be adopted, which can be fixed by the Dirac-Brueckner calculations of nuclear matter.

The field tensors for $\omega$-meson, $\rho$-meson, and photons ($A_\mu$) are given by
\begin{eqnarray}
\omega_{\mu\nu} &=& \partial_\mu \omega_\nu - \partial_\nu \omega_\mu, \\
\boldsymbol{\rho}_{\mu\nu}
  &=& \partial_\mu \boldsymbol{\rho}_\nu - \partial_\nu \boldsymbol{\rho}_\mu, \\
A_{\mu\nu} &=& \partial_\mu A_\nu - \partial_\nu A_\mu.
\end{eqnarray}
Then the equations of motion for fermions and bosons are obtained based on the Euler-Lagrange equation. For a system with time-reversal symmetry, the space-like components of the vector fields $\omega_\mu$ and $\boldsymbol{\rho}_\mu$ vanish, while charge conservation guarantees that only the 3rd component in the isospin space of $\boldsymbol{\rho}$ meson survives. In the mean field approximation (MFA), the meson fields become their mean values ($\sigma$, $\omega_0$, and $\rho_{0,3}$). The Klein-Gordon equations for bosons under MFA become
\begin{eqnarray}
(-\nabla^2 + m_\sigma^2) \sigma &=& - g_{\sigma} n_\mathrm{s} - U'(\sigma), \label{eq:KG_sigma} \\
(-\nabla^2 + m_\omega^2) \omega_0 &=& g_{\omega} n_\mathrm{b} - 2 \Lambda_\mathrm{v}g_\omega^2 g_\rho^2 \omega_0 \rho_{0,3}^2, \label{eq:KG_omega}\\
(-\nabla^2 + m_\rho^2) \rho_{0,3} &=&  \sum_{i=n,p} g_{\rho}\tau_{i, 3} n_i - 2 \Lambda_\mathrm{v}g_\omega^2 g_\rho^2 \omega_0^2\rho_{0,3}, \label{eq:KG_rho}\\
- \nabla^2 A_0 &=& e n_p - e n_e. \label{eq:KG_photon}
\end{eqnarray}
Here the nucleon scalar and vector densities are obtained with $n_\mathrm{s} = \sum_{i=n,p}\langle \bar{\psi}_i\psi_i \rangle$ and $n_\mathrm{b} = \sum_{i=n,p} n_i = \sum_{i=n,p}\langle \bar{\psi}_i\gamma^0 \psi_i \rangle$.

Since we are working with systems comprised of large numbers of particles, it is convenient to adopt the Thomas-Fermi approximation for fermions, where $\psi_i$ is considered as plane waves and the eigenvalues of the Dirac Equations are
\begin{eqnarray}
\epsilon_i^\pm(p) &=& g_{i\omega} \omega_0 + g_{i\rho}\tau_{i, 3} \rho_{0, 3} + q_i  A_0
              \pm \sqrt{p^2+{m_i^*}^2}, \label{eq:eigen}
\end{eqnarray}
with the effective nucleon mass being $m_n^*=m_p^*=m_N^*\equiv m_N + g_{\sigma} \sigma$ and $m_e^*=m_e = 0.511$ MeV.

The total energy of the system is obtained with
\begin{equation}
E=\int \langle {\cal{T}}_{00} \rangle \mbox{d}^3 r, \label{eq:energy}
\end{equation}
where the energy momentum tensor at zero temperature can be estimated with
\begin{eqnarray}
\langle {\cal{T}}_{00} \rangle
&=& \mathcal{E}_0
     + \frac{1}{2}(\nabla \sigma)^2 + \frac{1}{2}m_\sigma^2 \sigma^2 + U(\sigma)
     + \frac{1}{2}(\nabla \omega_0)^2  \nonumber \\
&&   + \frac{1}{2}m_\omega^2 \omega_0^2
     + \frac{1}{2}(\nabla \rho_{0,3})^2 + \frac{1}{2}m_\rho^2 \rho_{0,3}^2 \nonumber \\
&&   + 3 \Lambda_\mathrm{v}g_\omega^2 g_\rho^2 \omega_0^2 \rho_{0,3}^2
     + \frac{1}{2}(\nabla A_0)^2.
\label{eq:ener_dens}
\end{eqnarray}
Adopting no-sea approximation, the local kinetic energy density is determined by
\begin{eqnarray}
\mathcal{E}_0 &=& \sum_{i=n,p} \int_0^{\nu_i} \frac{p^2}{\pi^2} \sqrt{p^2+{m^*_N}^2}\mbox{d}p + \int_0^{\nu_e} \frac{p^2}{\pi^2} \sqrt{p^2+{m_e}^2}\mbox{d}p, \nonumber\\
              &=& \sum_{i=n,p} \frac {{m^*_N}^4}{8\pi^{2}}f\left(\frac{\nu_i}{m^*_N}\right) + \frac {m_e^4}{8\pi^{2}}f\left(\frac{\nu_e}{m_e}\right),
\end{eqnarray}
where $f(x) = \left[x(2x^2+1)\sqrt{x^2+1}-\mathrm{arcsh}(x) \right]$, and $\nu_i$ is the Fermi momentum and corresponds to the top of Fermi-sea, i.e., $\epsilon_i^+(\nu_i) = \mu_i = \rm{constant}$ with $\mu_i$ being the chemical potential. The source currents at zero temperature can be obtained with
\begin{eqnarray}
n_{s} &=& \sum_{i=n,p} \langle \bar{\psi}_i \psi_i \rangle = \sum_{i=n,p} \frac{{m^*_N}^3}{2\pi^2} g\left(\frac{\nu_i}{m^*_N}\right),\\
n_i &=& \langle \bar{\psi}_i\gamma^0 \psi_i \rangle = \frac{\nu_i^3}{3\pi^2},
\end{eqnarray}
where $g(x) = x \sqrt{x^2+1} - \mathrm{arcsh}(x)$. Note that $n_{s}$, $n_i$, and $\mathcal{E}_0$ represent the local properties of nuclear matter and vary with the space coordinates, which can be determined by the constancy of the chemical potentials, i.e.,
\begin{eqnarray}
\mu_i(\vec{r}) &=& g_{\omega} \omega_0(\vec{r}) + g_{\rho}\tau_{i, 3} \rho_{0, 3}(\vec{r}) + q_i  A_0(\vec{r}) \nonumber \\
&&  + \sqrt{{\nu_i(\vec{r})}^2+{m_i^*(\vec{r})}^2}  = \rm{constant}. \label{eq:chem_cons}
\end{eqnarray}

\section{\label{sec:num} Numerical details}

\subsection{\label{sec:num_pasta} Nuclear pasta}
In order to obtain the nonuniform structures of nuclear pasta, we need to solve the Klein-Gordon equations~(\ref{eq:KG_sigma}-\ref{eq:KG_photon}) and density distributions of fermions with Eq.~(\ref{eq:chem_cons}) based on mean field and Thomas-Fermi approximations. In previous investigations~\cite{Okamoto2012_PLB713-284, Okamoto2013_PRC88-025801}, Eqs.~(\ref{eq:KG_sigma}-\ref{eq:KG_photon}) and~(\ref{eq:chem_cons}) were solved iteratively inside a 3D periodic cell with discretized space coordinates. The body-centered cubic (BCC) and face-centered cubic (FCC) lattices for droplets/bubbles, simple and honeycomb configurations for rods/tubes, and only one type of slabs are found to be more stable than other exotic structures, which are symmetric under reflection. In such cases, to improve the computational efficiency, we expand the mean fields ($\phi=\sigma$, $\omega_0$, $\rho_{0,3}$, $A_0$) as
\begin{equation}
 \phi(\vec{r}) = \sum_{i,j,k} \tilde{\phi}_{i,j,k} \cos(x p_{xi})\cos(y p_{yj})\cos(z p_{zk}), \label{eq:phi}
\end{equation}
which is equivalent to consider one octant of the unit cell~\cite{Newton2009_PRC79-055801}. In principle, we can further reduce the computational cost by considering one octant of the WS cell. We did not do so in order to include the possible emergency of intermediate structures~\cite{Watanabe2003_PRC68-035806, Okamoto2012_PLB713-284}. The indices $i$, $j$, $k$ run from 0 to $N_{x,y,z}-1$, where $N_{x,y,z}$ is the total grid number on $x$-, $y$-, and $z$-axis, respectively. The quantities $p_{xi}$, $p_{yj}$, and $p_{zk}$ take discrete values and are determined by
\begin{equation}
 p_{xi} = \frac{\pi i}{\Delta x N_x}, \ p_{yj} = \frac{\pi j}{\Delta y N_y}, \ p_{zk} = \frac{\pi k}{\Delta z N_z},
\end{equation}
where $\Delta x$, $\Delta y$, and $\Delta z$ are the grid distances on $x$-, $y$-, and $z$-axis. The space coordinates in Eq.~(\ref{eq:phi}) thus lie within $-\Delta x N_x \leq x \leq \Delta x N_x$, $-\Delta y N_y \leq y \leq \Delta y N_y$, and $-\Delta z N_z \leq z \leq \Delta z N_z$. The coefficients $\tilde{\phi}_{i,j,k}$ are fixed by solving the Klein-Gordon equations~(\ref{eq:KG_sigma}-\ref{eq:KG_photon}), which are now reduced to
\begin{equation}
 \tilde{\phi}_{i,j,k} = \frac{S_{i,j,k}}{p_{xi}^2 + p_{yj}^2 + p_{zk}^2 + m_\phi^2},  \label{eq:phi_coeff}
\end{equation}
with the source currents $S_{i,j,k}$ obtained via fast cosine transformations on the right hand sides of Eqs.~(\ref{eq:KG_sigma}-\ref{eq:KG_photon}). Based on Eq.~(\ref{eq:phi}), the energy contributions of the terms $\frac{1}{2} \int (\nabla \phi)^2 \mbox{d}^3 r $ in Eq.~(\ref{eq:energy}) are determined with these coefficients, i.e.,
\begin{equation}
 \int (\nabla \phi)^2 \mbox{d}^3 r =  \sum_{i,j,k} \frac{V}{h_i h_j h_k}\tilde{\phi}_{i,j,k}^2 \left(p_{xi}^2 + p_{yj}^2 + p_{zk}^2 \right),
\end{equation}
where the volume $V =\Delta x\Delta y\Delta z N_x N_y N_z$ takes one octant of the unit cell, the coefficients $h_0 = 1$ and $h_{i} = 2$ at $i>0$. Once we obtain the mean fields with Eq.~(\ref{eq:phi}), the local chemical potentials are determined by Eq.~(\ref{eq:chem_cons}). To reach the ground state, the density distributions of nucleons and electrons should meet the requirement of the constancy of chemical potentials. In practice, in order to fulfill Eq.~(\ref{eq:chem_cons}), we adopt the imaginary time step method~\cite{Levit1984_PLB139-147} and solve Eqs.~(\ref{eq:KG_sigma}-\ref{eq:KG_photon}) and~(\ref{eq:chem_cons}) iteratively. In summary, Eqs.~(\ref{eq:KG_sigma}-\ref{eq:KG_photon}) and~(\ref{eq:chem_cons}) are solved iteratively inside a 3D periodic unit cell with discretized space coordinate and reflection symmetry, i.e.,
\begin{enumerate}
  \item \label{item:itr_1} Assume initial density distributions of fermions at given total particle numbers;
  \item \label{item:itr_2} Solve the Klein-Gordon equations~(\ref{eq:KG_sigma}-\ref{eq:KG_photon}) with Eq.~(\ref{eq:phi_coeff}) via fast cosine transformations;
  \item \label{item:itr_3} Obtain the local chemical potentials with Eq.~(\ref{eq:chem_cons}) according to the mean fields determined by Eq.~(\ref{eq:phi});
  \item \label{item:itr_4} Readjust the density distributions of fermions with the imaginary time step method~\cite{Levit1984_PLB139-147};
  \item \label{item:itr_5} Go to step~\ref{item:itr_2} until convergence is reached;
  \item \label{item:itr_6} Obtain the energy of the system with Eq.~(\ref{eq:energy}).
\end{enumerate}
By properly choosing the initial density profiles, the pasta structure will eventually evolve into certain configurations via imaginary time step method. If a random initial density profile was applied~\cite{Okamoto2012_PLB713-284, Okamoto2013_PRC88-025801}, we have little control over the converged pasta structure. It is thus more efficient to assume some initial configurations, which will normally evolve into the chosen nuclear pasta structure. The ground state structure can then be obtained by searching for the configuration that gives the minimum energy per baryon.

\begin{figure}
\includegraphics[width=\linewidth]{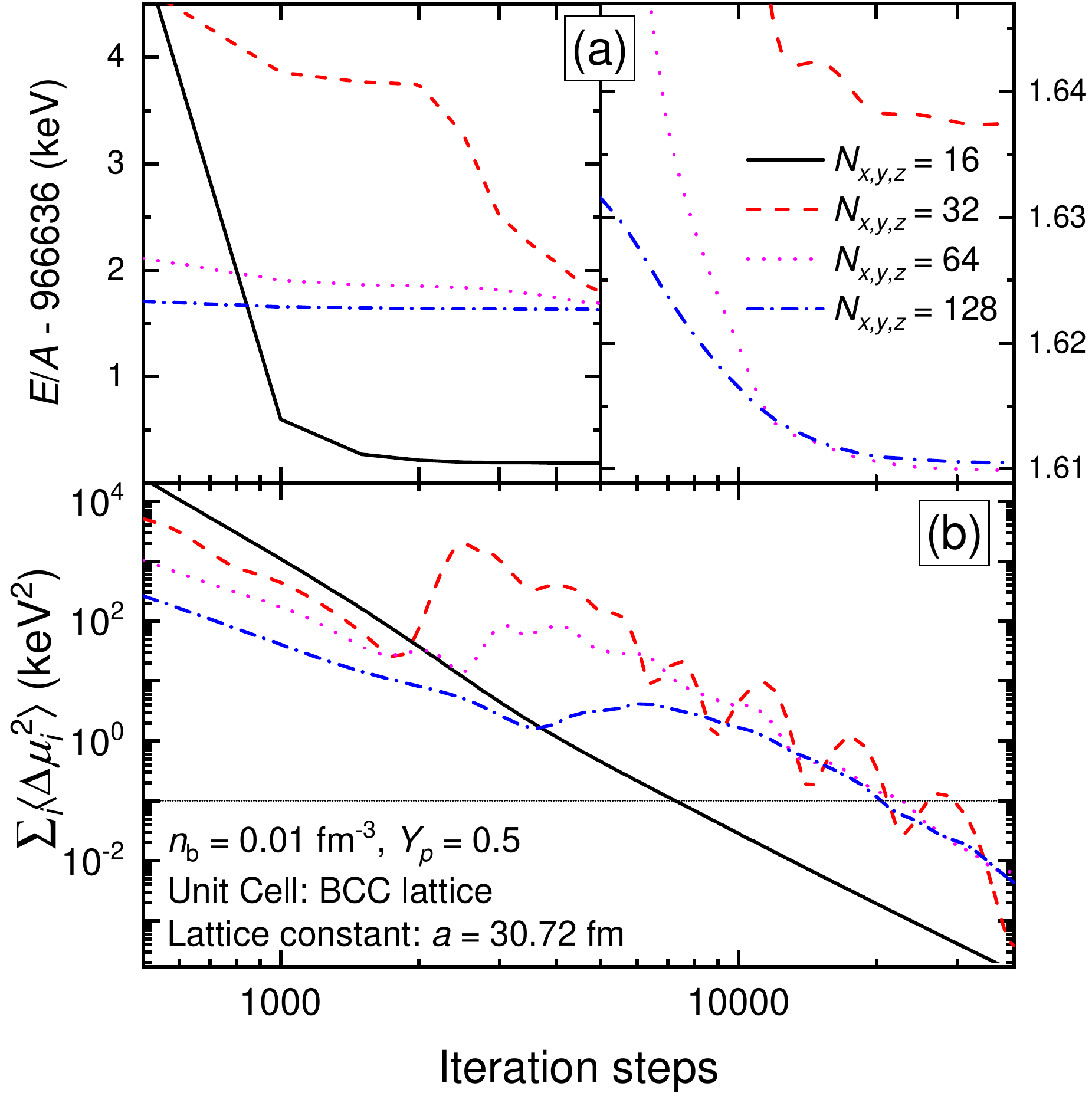}
\caption{\label{Fig:Converge} Energy per baryon (a) of nuclear pasta in BCC lattice and total deviation (b) of local chemical potentials as functions of iteration steps.}
\end{figure}

To check if Eq.~(\ref{eq:chem_cons}) is satisfied, we examine the deviation of local chemical potentials, i.e.,
\begin{equation}
 \sum_{i=p,n,e} \langle \Delta \mu_i^2 \rangle = \sum_{i=p,n,e} \frac{ \int \left[\mu_i(\vec{r}) - \bar{\mu}_i\right]^2 n_i(\vec{r}) \mbox{d}^3 r} {\int n_i(\vec{r}) \mbox{d}^3 r}, \label{eq:converg}
\end{equation}
where $\bar{\mu}_i = \left.\int \mu_i(\vec{r}) n_i(\vec{r}) \mbox{d}^3 r\right/\int n_i(\vec{r}) \mbox{d}^3 r$ is the average chemical potential. As an example, in Fig.~\ref{Fig:Converge} we present the obtained energy per baryon of nuclear pasta in a BCC lattice with the lattice constant $a = 30.72$ fm, where the parameter set (Set 0 in Tab.~\ref{tab:NM}) introduced in Ref.~\cite{Maruyama2005_PRC72-015802} is adopted. The corresponding grid distances are obtained with $\Delta x = \Delta y = \Delta z = a/(2N_{x,y,z})$. If we set $\sum_{i=p,n,e} \langle \Delta \mu_i^2 \rangle <0.1\ \mathrm{keV}^2$ as the convergency condition, the deviation of energy per baryon from the fully converged result is found to be $\Delta E/A\lesssim 0.001$ keV. Meanwhile, as indicated in Fig.~\ref{Fig:Converge}, varying the grid distance will cause larger deviations compared with the energy per baryon $E/A$ obtained at $N_{x,y,z} = 128$, which are $\Delta E/A \approx 1.5$ keV, 0.03 keV, and 0.0006 keV for $\Delta x = \Delta y = \Delta z = 0.96$ fm, 0.48 fm, and 0.24 fm, respectively. If less grid points and larger grid distances are adopted, we expect larger deviations on energy per baryon.

\begin{figure}
\includegraphics[width=\linewidth]{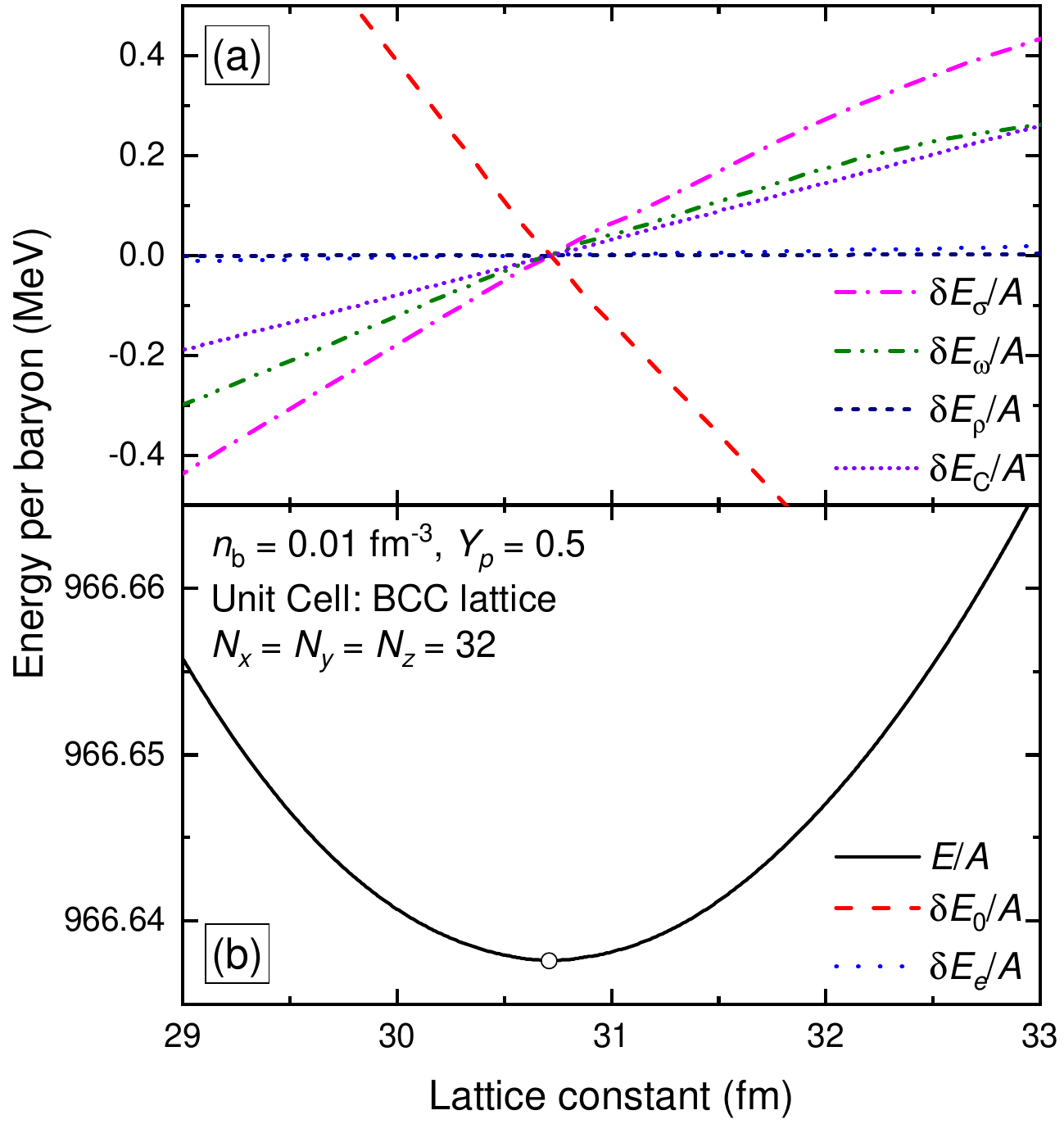}
\caption{\label{Fig:Box} Relative energies (a) and energy (b) per baryon of nuclear pasta in BCC lattice as functions of the lattice constant $a$, corresponding to Fig.~\ref{Fig:Converge}. The open circle indicates the optimal lattice constant at $a = 30.71$ fm.}
\end{figure}

Besides the issue of convergency, another factor that affects our prediction is the effect of finite cell size~\cite{GimenezMolinelli2014_NPA923-31, Newton2009_PRC79-055801}. Since we are working with only one octant of the unit cell, the obtained energy per baryon and pasta structure are sensitive to the lattice constants $a$, $b$, and $c$. We thus vary the lattice constants and search for the minimum $E/A$ at fixed nuclear shape, lattice structure, baryon number density $n_\mathrm{b}  \equiv \int \left[n_p(\vec{r}) + n_n(\vec{r})\right]\mbox{d}^3 r / V$, and proton fraction $Y_p  \equiv \int n_p(\vec{r}) \mbox{d}^3 r / (V n_\mathrm{b})$. As an example, in Fig.~\ref{Fig:Box} we present the obtained energies per baryon of nuclear pasta in BCC lattice as functions of the lattice constant $a$ ($=b=c$). The kinetic energy $E_0$, kinetic energy of electrons $E_e$, energy contributions from $\sigma$-, $\omega$-, $\rho$-mesons $E_{\sigma, \omega, \rho}$, and Coulomb energy $E_\mathrm{C}$ are obtained with
\begin{eqnarray}
E_0 &=& \int \mathcal{E}_0 \mbox{d}^3 r,\\
E_e &=& \frac{m_e^4}{8\pi^{2}} \int f\left(\frac{\nu_e}{m_e}\right) \mbox{d}^3 r,\\
E_\sigma &=& \int \left[ \frac{1}{2}(\nabla \sigma)^2 + \frac{1}{2}m_\sigma^2 \sigma^2 + U(\sigma) \right] \mbox{d}^3 r,\\
E_\omega &=& \frac{1}{2}\int \left[ (\nabla \omega_0)^2 + m_\omega^2 \omega^2  \right] \mbox{d}^3 r,\\
E_\rho &=&  \frac{1}{2}\int \left[ (\nabla \rho_{0,3})^2 + m_\rho^2 \rho_{0,3}^2  \right] \mbox{d}^3 r,\\
E_\mathrm{C} &=&  \frac{1}{2}\int (\nabla A_0)^2\mbox{d}^3 r,
\end{eqnarray}
where their relative values with respect to those at the optimal lattice constant ($a = 30.71$ fm) are shown in Fig.~\ref{Fig:Box} (b). As we increase the lattice constant (box size), the sizes of droplets increase as well, which reduces the surface energy per baryon as $\left(E_0 + E_\sigma + E_\omega + E_\rho - E_e\right)/A$ decreases. Meanwhile, the Coulomb energy per baryon $E_\mathrm{C}/A$ increases almost linearly with $a$. A balance between the energy contributions from the nuclear part and Coulomb part is then attained at the optimal lattice constant $a = 30.71$ fm. Note that electrons have little impact on the optimal size of unit cell, while the contribution of $\rho$-mesons $E_{\rho}$ is insignificant since we are considering only the symmetric nuclear matter. At fixed nuclear shape, lattice structure, baryon number density $n_\mathrm{b}$, and proton fraction $Y_p$, we then carry out similar procedures to determine the optimal lattice constants and minimum energy.

Finally, we have examined multiple unit cells by expanding the obtained one octant of the unit cell, the deviation of energy per baryon lies within the uncertainty range corresponding to the convergency condition $\sum_{i=p,n,e} \langle \Delta \mu_i^2 \rangle <0.1\ \mathrm{keV}^2$. We thus search for the ground state configurations considering only one octant of the unit cell, where the simple cubic (SC), BCC, and FCC lattices for droplets/bubbles, simple and honeycomb configurations for rods/tubes, and slabs are examined at various combinations of $n_\mathrm{b}$ and $Y_p$.

\subsection{\label{sec:num_sym} Symmetry energy and $\omega$-$\rho$ coupling}
For the isoscalar channel of the effective $N$-$N$ interactions in RMF model, we adopt the parameter set proposed in Ref.~\cite{Maruyama2005_PRC72-015802}. According to Ref.~\cite{Maruyama2005_PRC72-015802}, the masses of nucleons $m_N = 938\ \rm{MeV}$, $\sigma$-mesons $m_\sigma = 400\ \rm{MeV}$, $\omega$-mesons $m_\omega = 783\ \rm{MeV}$, and $\rho$-mesons $m_\sigma = 769\ \rm{MeV}$. The nucleon-meson coupling constants $g_\sigma = 6.3935$  and $g_{\omega} = 8.7207$, while the coefficients of the nonlinear self-couplings of $\sigma$ are $b= -0.008659$ and $c = -0.002421$. These parameters are fixed to reproduce the properties of nuclear matter at the saturation density $n_0 = 0.153\ \rm{fm}^{-3}$, i.e., the binding energy per baryon $B(n_0) = E/A - m_N = -16.3$ MeV, the incompressibility $K(n_0) = 240$ MeV, and the effective nucleon mass $m_N^*(n_0) = 0.78 m_N$.

\begin{figure}
\includegraphics[width=\linewidth]{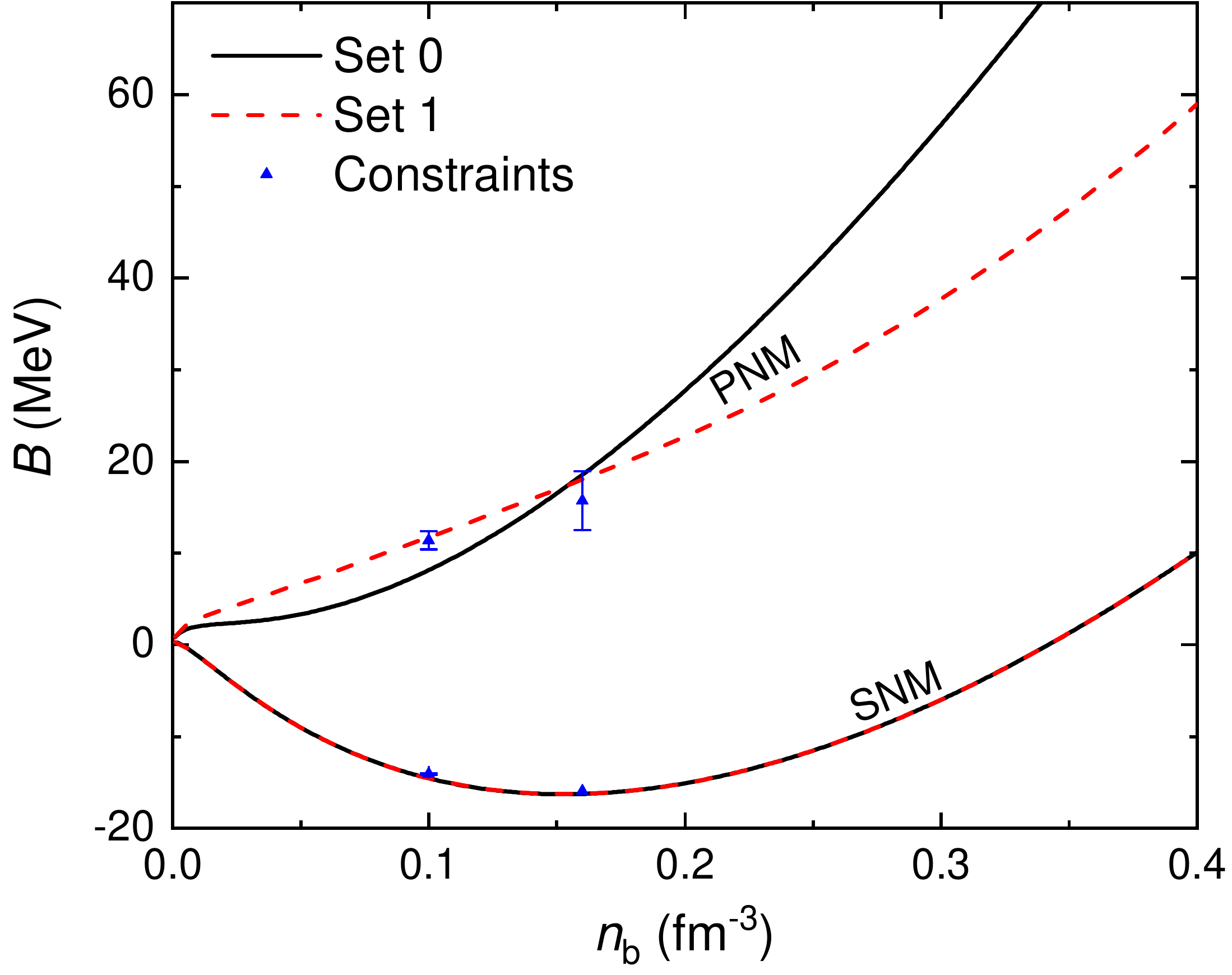}
\caption{\label{Fig:EpA_Uniform} Binding energy per nucleon for symmetric nuclear matter (SNM) and pure neutron matter (PNM) predicted by two sets of parameters in Tab.~\ref{tab:NM}. The constrains  $B_\mathrm{PNM}(n_\mathrm{on}) = 11.4\pm1.0$ MeV, $B_\mathrm{SNM}(n_\mathrm{on}) = -14.1\pm0.1$ MeV~\cite{Brown2013_PRL111-232502}, $B_\mathrm{SNM}(n_0) = -16$ MeV, and $B_\mathrm{PNM}(n_0) = B_\mathrm{SNM}(n_0) + S(n_0) = 15.7 \pm 3.2$ MeV~\cite{Li2013_PLB727-276, Oertel2017_RMP89-015007} are indicated with solid triangles.}
\end{figure}

\begin{table}
\caption{\label{tab:NM} The symmetry energy and its slope of nuclear matter, neutron skin thickness of $^{208}$Pb, maximum masses $M_\mathrm{max}$, radii $R_{1.4}$ and tidal deformation of 1.4 solar-mass neutron stars predicted by two sets of parameters (0 and 1) in the isovector channel. Here Set 0 takes the same parameters as in Ref.~\cite{Maruyama2005_PRC72-015802}, while the updated Set 1 gives a smaller slope of symmetry energy to coincide with the recent astrophysical and chiral EFT constraints~\cite{Essick2021}. For both Set 0 and Set 1, the parameters proposed in Ref.~\cite{Maruyama2005_PRC72-015802} are adopted in the isoscalar channel. }
\begin{tabular}{c|cc|cc|cccc} \hline \hline
      &$g_{\rho}$& $\Lambda_\mathrm{v}$ & $S$   & $L$   & $\Delta R_{np}$ &$M_\mathrm{max}$ & $R_{1.4}$ & $\Lambda_{1.4}$   \\
      &          &                      &  MeV  &  MeV  &      fm         &$M_\odot$        & km        &             \\   \hline
Set 0 & 4.2696   & 0                    & 32.46 & 89.39 &    0.195        & 2.02            & 13.1      & 624        \\
Set 1 & 5.55048  & 0.34                 & 31.85 & 41.34 &    0.157        & 1.98            & 11.9      & 331       \\
\hline
\end{tabular}
\end{table}

For the isovector channel, we consider two scenarios as indicated in Tab.~\ref{tab:NM}. Set 0 corresponds to the original cases in Ref.~\cite{Maruyama2005_PRC72-015802}, which predicts the symmetry energy $S(n_0) = 32.46$ MeV and its slope $L(n_0) = 89.39$ MeV. According to the recent constraints on the tidal deformability $70\leq \Lambda_{1.4}\leq 580$ from the GW170817 binary neutron star merger event~\cite{LVC2018_PRL121-161101}, a smaller slope of symmetry energy is preferred~\cite{Zhu2018_ApJ862-98, Tsang2019_PLB795-533, Dexheimer2019_JPG46-034002, Zhang2019_EPJA55-39, Zhang2020_PRC101-034303, Li2020_PRC102-045807}. In such cases, we adopt an $\omega$-$\rho$ cross coupling term $\Lambda_\mathrm{v}g_\omega^2 g_\rho^2 (\omega_\mu\omega^\mu) (\boldsymbol{\rho}_\mu\cdot\boldsymbol{\rho}^\mu)$ and reduce $L$ by readjusting $g_{\rho}$ and $\Lambda_\mathrm{v}$~\cite{Shen2020_ApJ891-148}. In practice, by keeping $S(n_0)$ within the range of $31.7 \pm 3.2$ MeV~\cite{Li2013_PLB727-276, Oertel2017_RMP89-015007},
we fix $g_{\rho}$ and $\Lambda_\mathrm{v}$ according to the binding energy of pure neutron matter (PNM) at $n_\mathrm{on} = 0.1\ \rm{fm}^{-3}$, where a robust constraint was found with $B_\mathrm{PNM}(n_\mathrm{on}) = 11.4\pm1.0$ MeV~\cite{Brown2013_PRL111-232502}. The new parameter set is then listed as Set 1 in Tab.~\ref{tab:NM}, which predicts a smaller slope of symmetry energy ($L=41.34$ MeV) compared with that of Set 0 ($L=89.39$ MeV) initially proposed in Ref.~\cite{Maruyama2005_PRC72-015802}. As will be addressed later, the slope of symmetry energy of Set 0 coincides with the recent measurement of PREX-II~\cite{PREX2021}, while that of Set 1 is consistent with various astrophysical and chiral effective field theory (EFT) constraints~\cite{Essick2021}. The obtained energy per baryon for both PNM and symmetric nuclear matter (SNM) are then presented in Fig.~\ref{Fig:EpA_Uniform}. Note that for SNM, the corresponding binding energy at $n_\mathrm{on} = 0.1\ \rm{fm}^{-3}$ is $B_\mathrm{SNM}(n_\mathrm{on}) = -14.1\pm0.1$ MeV~\cite{Brown2013_PRL111-232502}, which is consistent with our predictions (Set 1) as well.

\begin{figure}
\includegraphics[width=\linewidth]{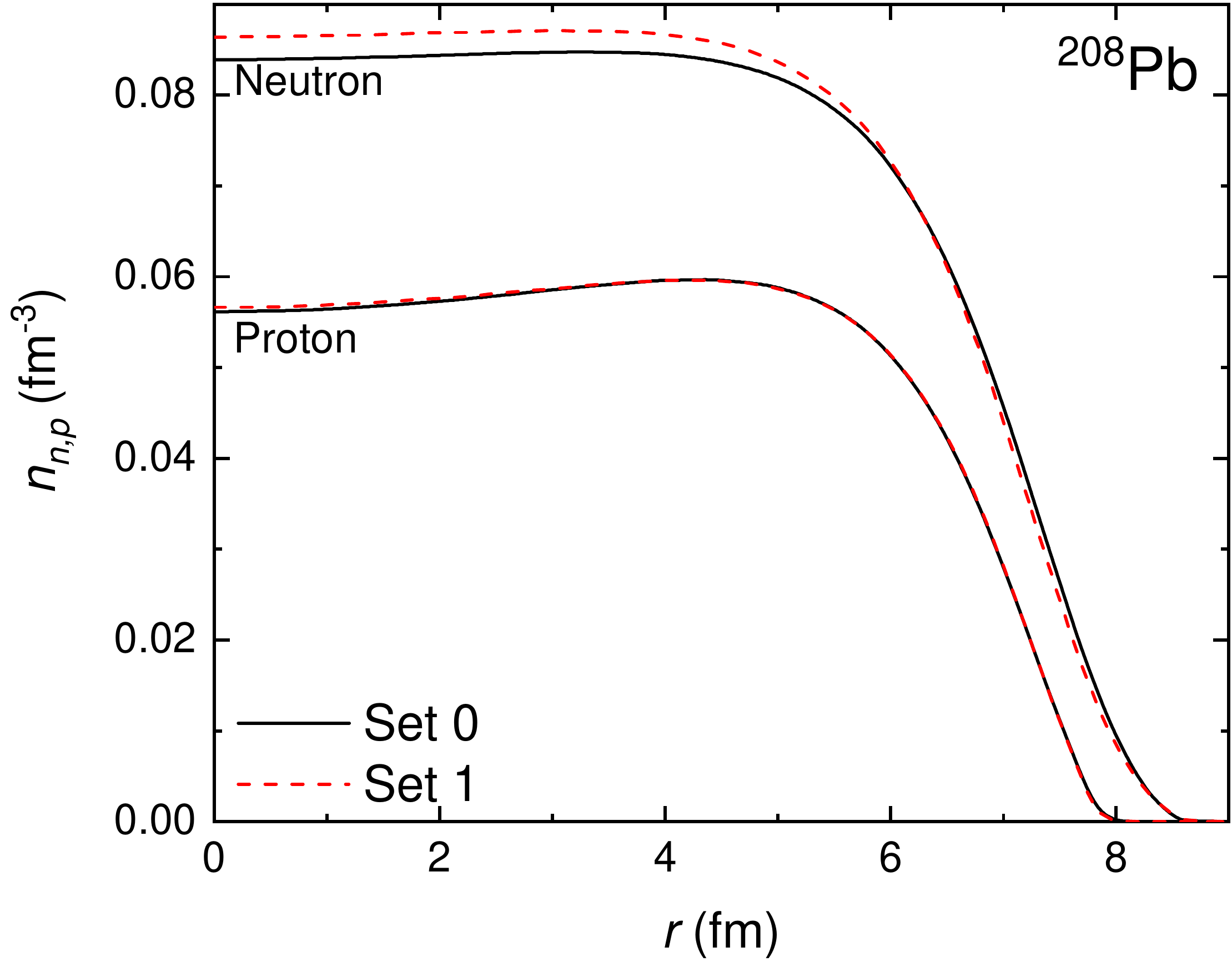}
\caption{\label{Fig:DensPb208} Density profiles of $^{208}$Pb obtained with Thomas-Fermi approximation using two sets of parameters in Tab.~\ref{tab:NM}.}
\end{figure}

In previous studies, it was shown that the slope of symmetry energy is sensitive to the neutron skin thickness and follows a linear correlation, e.g., $\Delta R_{np} = 0.101+0.00147 L$ for $^{208}$Pb~\cite{Brown2000_PRL85-5296, Roca-Maza2011_PRL106-252501}. In Fig.~\ref{Fig:DensPb208} we present the obtained density profiles of $^{208}$Pb in Thomas-Fermi approximation, where the parameter sets listed in Tab.~\ref{tab:NM} are adopted. We note that the proton density profiles are close to each other, while neutrons are more concentrated at the center for Set 1. This is mainly because Set 1 predicts larger symmetry energy at subsaturation densities, which provides stronger proton-neutron attractive interactions. Based on the density profiles in Fig.~\ref{Fig:DensPb208}, the neutron skin thickness of $^{208}$Pb can be estimated with
\begin{equation}
\Delta R_{np} = \sqrt{\langle r_n^2 \rangle} - \sqrt{\langle r_p^2 \rangle},
\end{equation}
where $\langle r_i^2 \rangle  = \int_0^\infty r^4 n_i(r) \mbox{d}r / \int_0^\infty r^2 n_i(r) \mbox{d}r$. The obtained $\Delta R_{np}$ corresponding to the two parameter sets are indicated in Tab.~\ref{tab:NM}, which lie within the experimental constraints $\Delta R_{np}=0.33^{+0.16}_{-0.18}$ fm measured in PREX-I~\cite{PREX2012_PRL108-112502}. A recent measurement with PREX-II suggests $\Delta R_{np}=0.283\pm 0.071$ fm~\cite{PREX2021}, which predicts a rather large slope of symmetry energy $L=106 \pm 37$ MeV. We find Set 0 with $L=89.39$ MeV lies within the range while Set 1 with $L=41.34$ MeV becomes too small. Note that the Thomas-Fermi approximation tends to underestimate the neutron skin thickness~\cite{Shen2020_ApJ891-148}, we thus expect slightly larger $\Delta R_{np}$ than those in Tab.~\ref{tab:NM}.


\begin{figure}
\includegraphics[width=\linewidth]{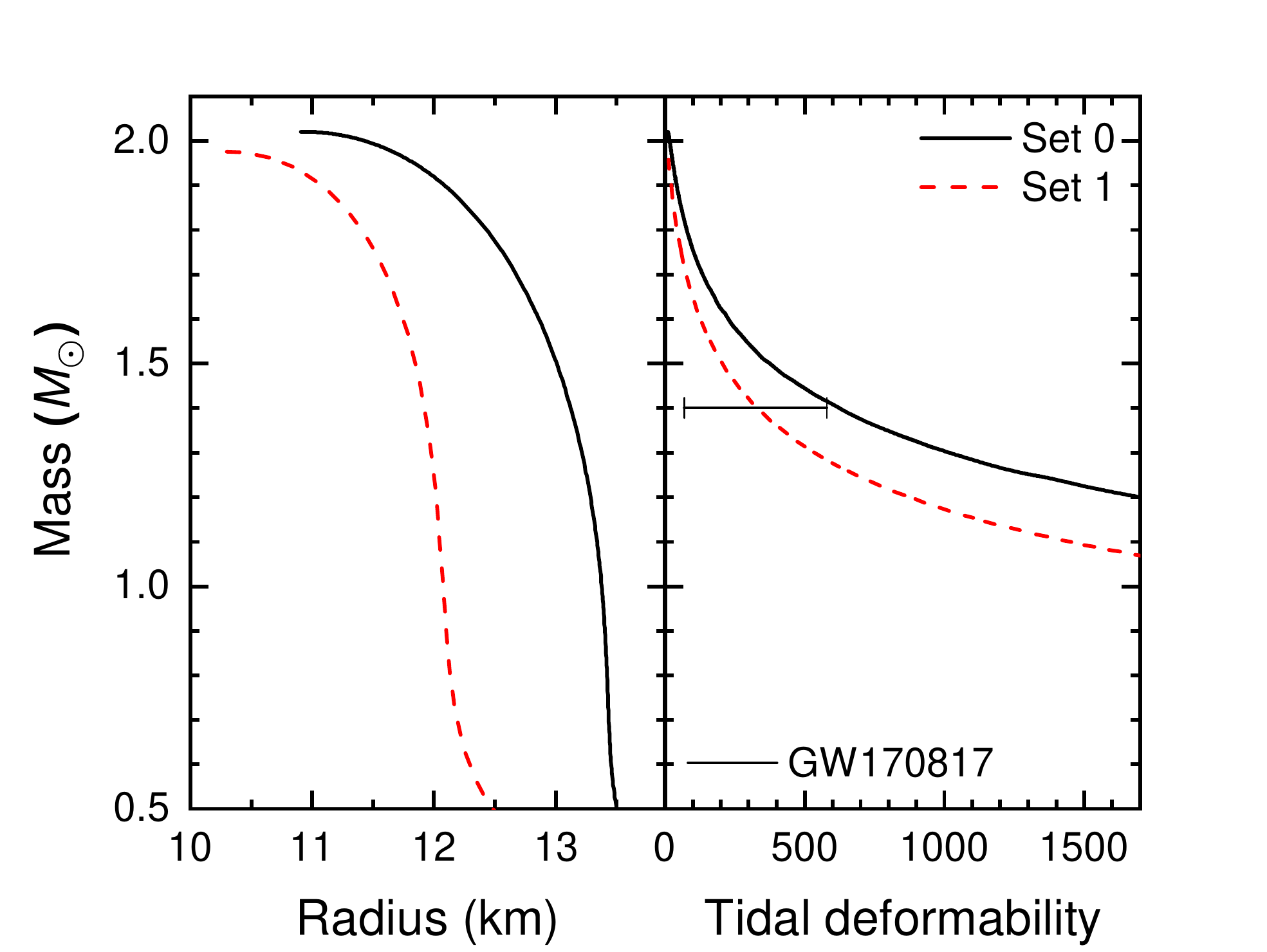}
\caption{\label{Fig:MRL} Mass, radius, and tidal deformability of neutron stars obtained with the two sets of parameters in Tab.~\ref{tab:NM}.}
\end{figure}

Based on the binary neutron star merger event GRB 170817A-GW170817-AT 2017gfo, more stringent constraint on $L$ can be obtained according to the measured tidal deformability of neutron stars~\cite{Zhu2018_ApJ862-98, Tsang2019_PLB795-533, Dexheimer2019_JPG46-034002, Zhang2019_EPJA55-39, Zhang2020_PRC101-034303, Li2020_PRC102-045807}. For $1.4 M_\odot$ neutron stars, its tidal deformability was constrained within $70\leq \Lambda_{1.4}\leq 580$~\cite{LVC2018_PRL121-161101}. In Fig.~\ref{Fig:MRL} we present the mass, radius, and tidal deformability of neutron stars predicted by the two sets of parameters in Tab.~\ref{tab:NM}, where the corresponding equation of states (EOSs) are plotted in Fig.~\ref{Fig:EpA_beta} with their numerical data indicated in \cite[Tab. II]{Maruyama2005_PRC72-015802} for Set 0 ($L=89.39$ MeV) and Tab.~\ref{table:EOS} for Set 1 ($L=41.34$ MeV). At $n_\mathrm{b}\leq 0.001\ \mathrm{fm}^{-3}$, we adopt the EOSs presented in Refs.~\cite{Feynman1949_PR75-1561, Baym1971_ApJ170-299, Negele1973_NPA207-298}. Evidently, the tidal deformability $\Lambda_{1.4}$ obtained with Set 0 exceeds the upper limit due to a larger $L$, while that of Set 1 is consistent with observation. The neutron stars' radii obtained with both parameter sets lie within the radius range (11.52-14.26 km) of PSR J0030+0451 measured in the NICER mission~\cite{Riley2019_ApJ887-L21, Miller2019_ApJ887-L24}. Note that the maximum masses of neutron stars are slightly smaller than the lower limit of the observational mass ($2.14^{+0.10}_{-0.09}M_{\odot}$) of PSR J0740+6620~\cite{Cromartie2020_NA4-72}. This is not a problem if we take the firmest limit of the 95.4\% confidence band, which reduces the lower limit to $1.96 M_{\odot}$~\cite{Cromartie2020_NA4-72} and permits both parameter sets. If exotic phases appear inside neutron stars, we expect the maximum masses to be further reduced. In such cases, our EOSs are not applicable at the center regions of massive neutron stars ($M\gtrsim 1.4 M_{\odot}$), where the density usually exceeds $\sim$$3 n_0$. At lower densities, however, our results should be valid, where smaller slopes of symmetry energy are favorable according to the constraints of $\Lambda_{1.4}$.

\section{\label{sec:pasta} Results and Discussion}

\begin{figure*}[htbp]
\hspace{-0.05\linewidth}
\begin{minipage}[t]{0.33\linewidth}
\includegraphics[width=1.23\textwidth]{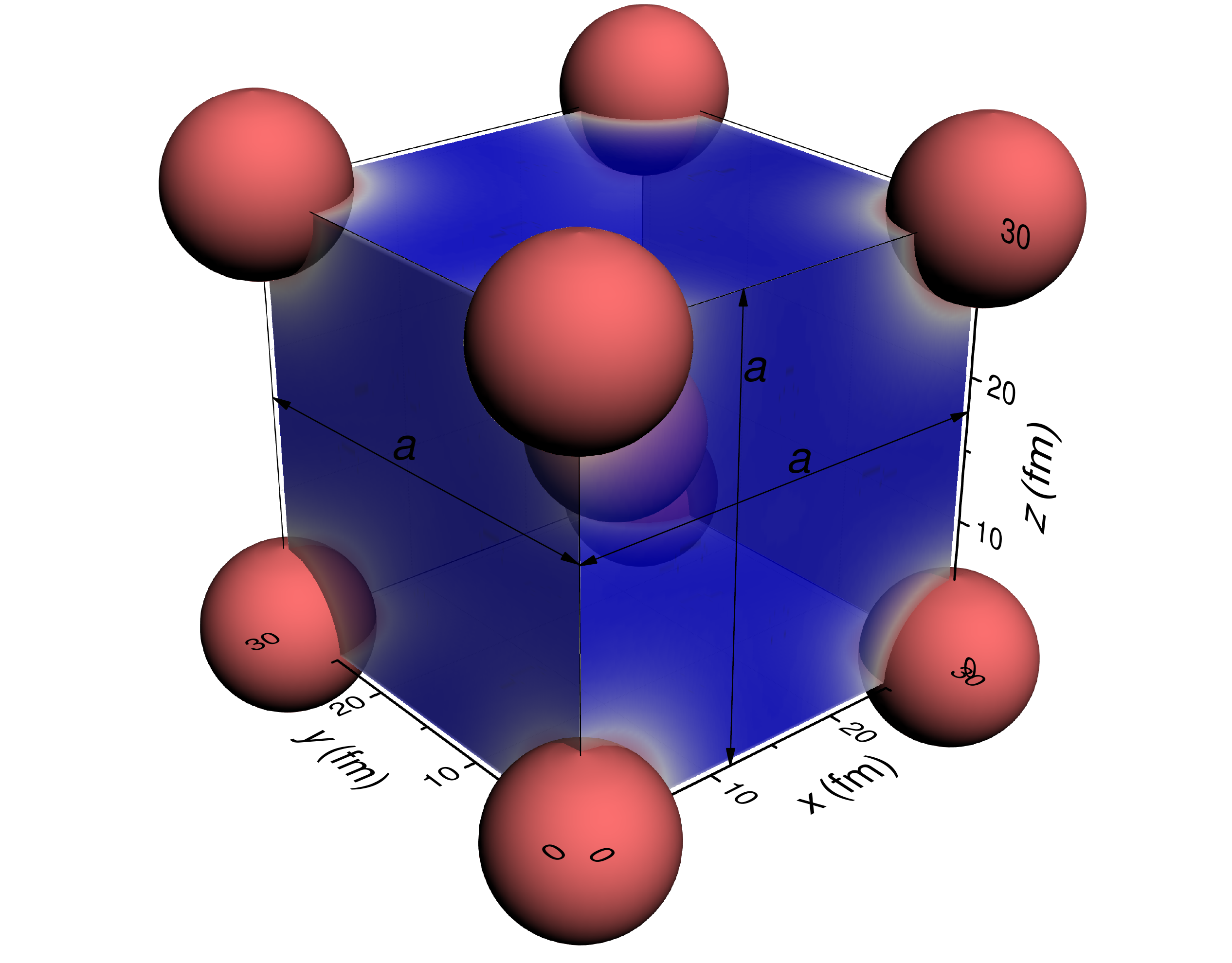}
(a)
\end{minipage}%
\hfill
\begin{minipage}[t]{0.33\linewidth}
\includegraphics[width=1.23\textwidth]{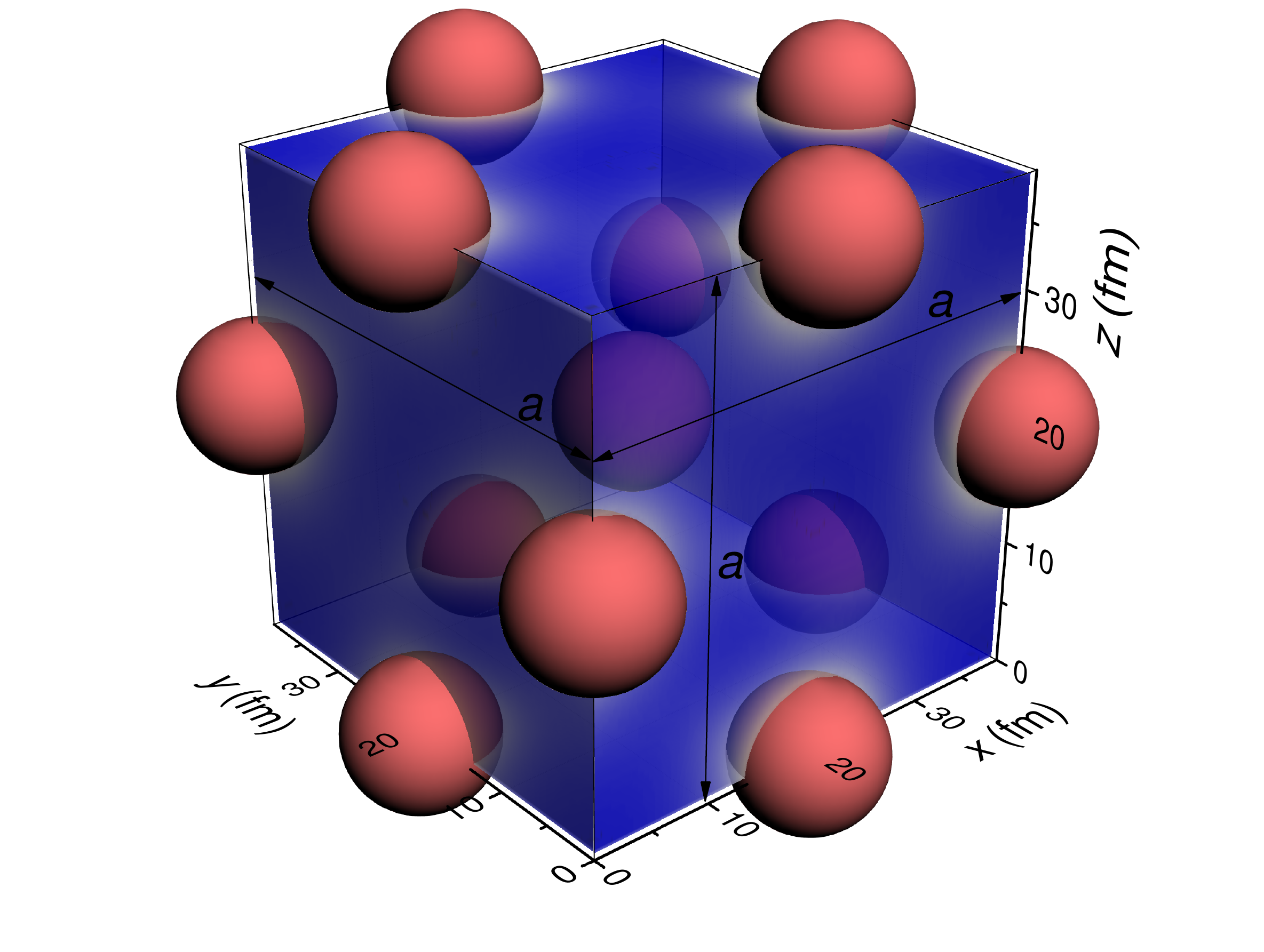}
(b)
\end{minipage}
\hfill
\begin{minipage}[t]{0.33\linewidth}
\includegraphics[width=1.23\textwidth]{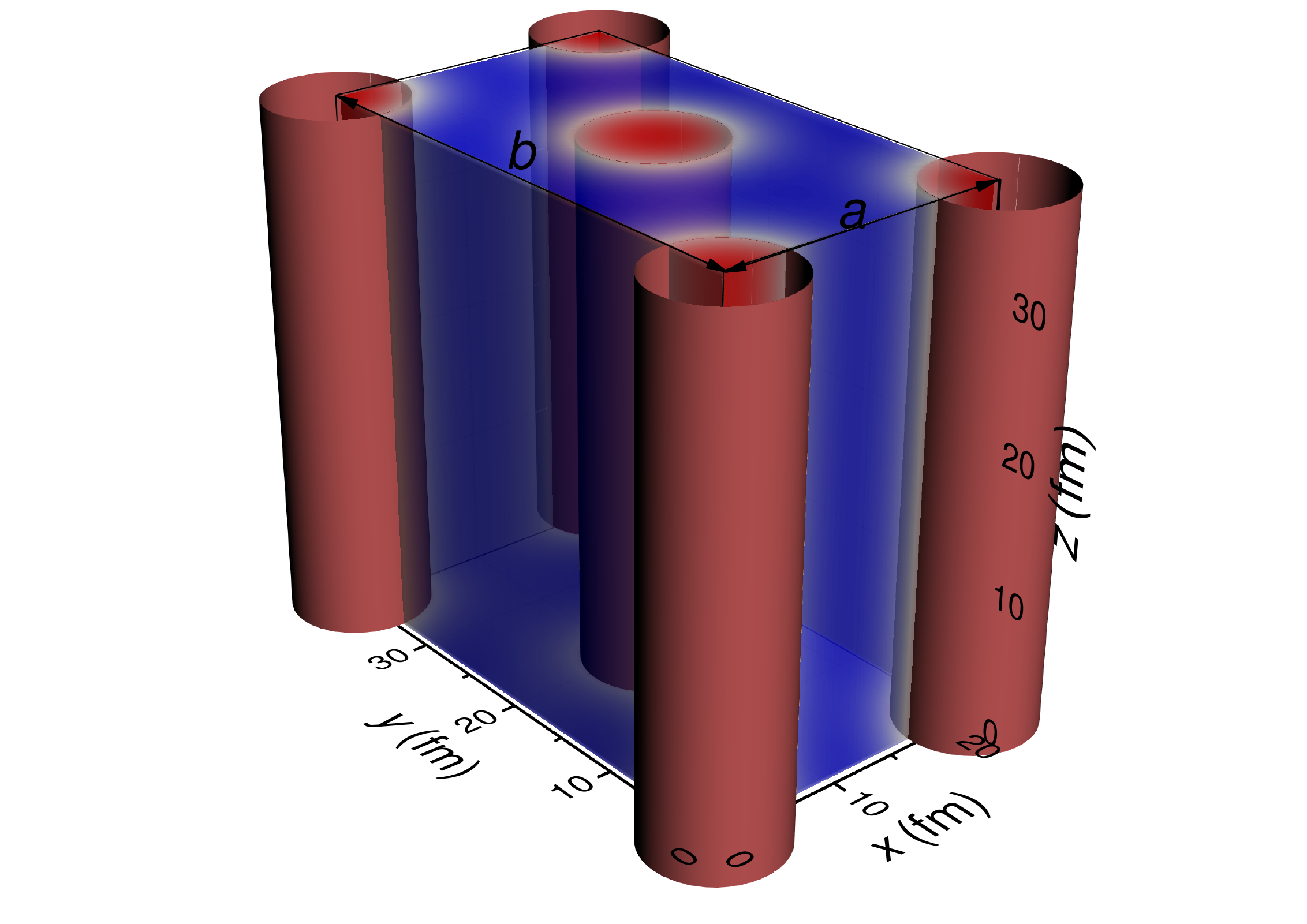}
(c)
\end{minipage}
\\
\begin{minipage}[t]{0.42\linewidth}
\includegraphics[width=\textwidth]{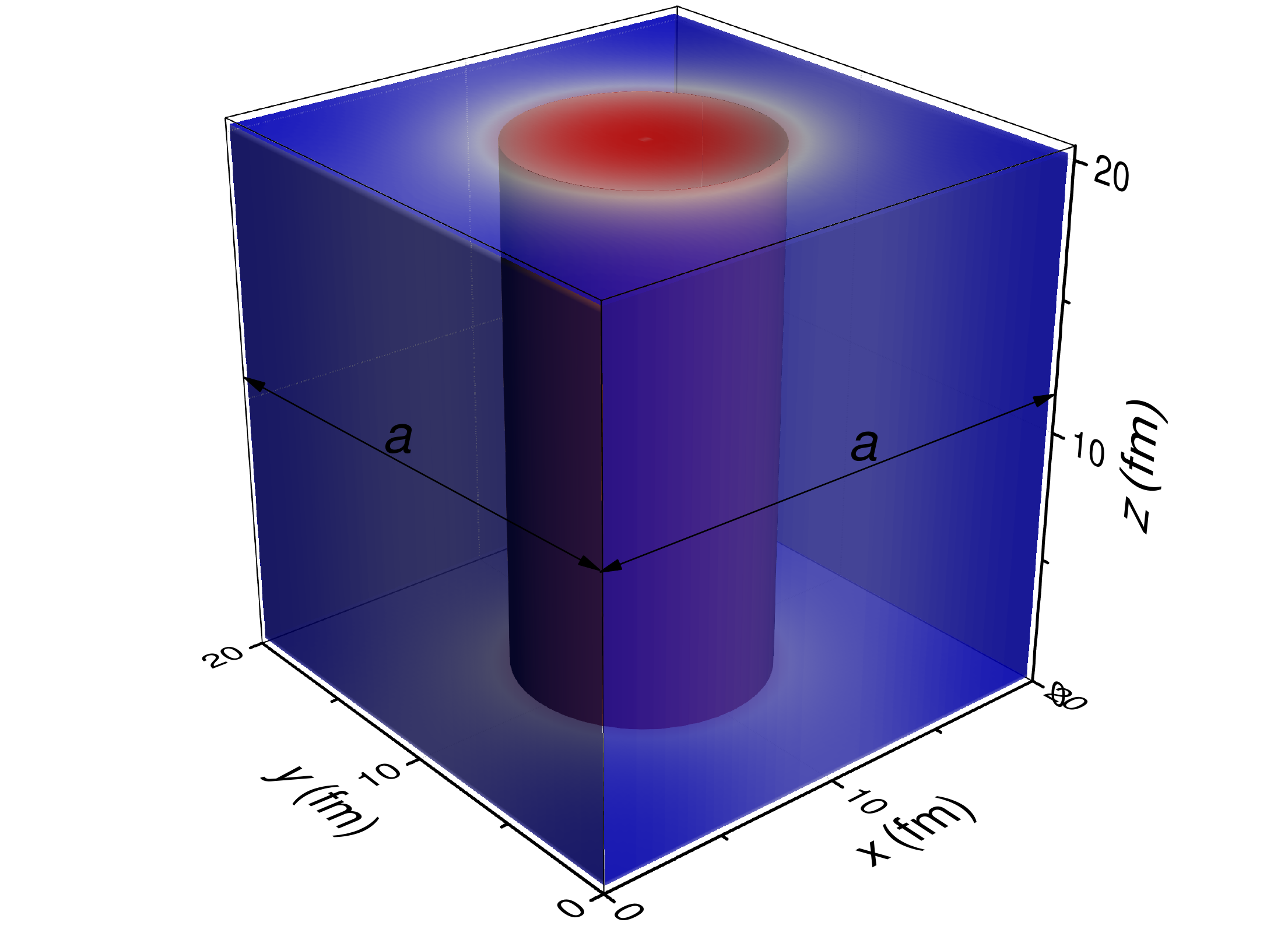}
(d)
\end{minipage}
\ \ \ \ \ \
\begin{minipage}[t]{0.42\linewidth}
\includegraphics[width=\textwidth]{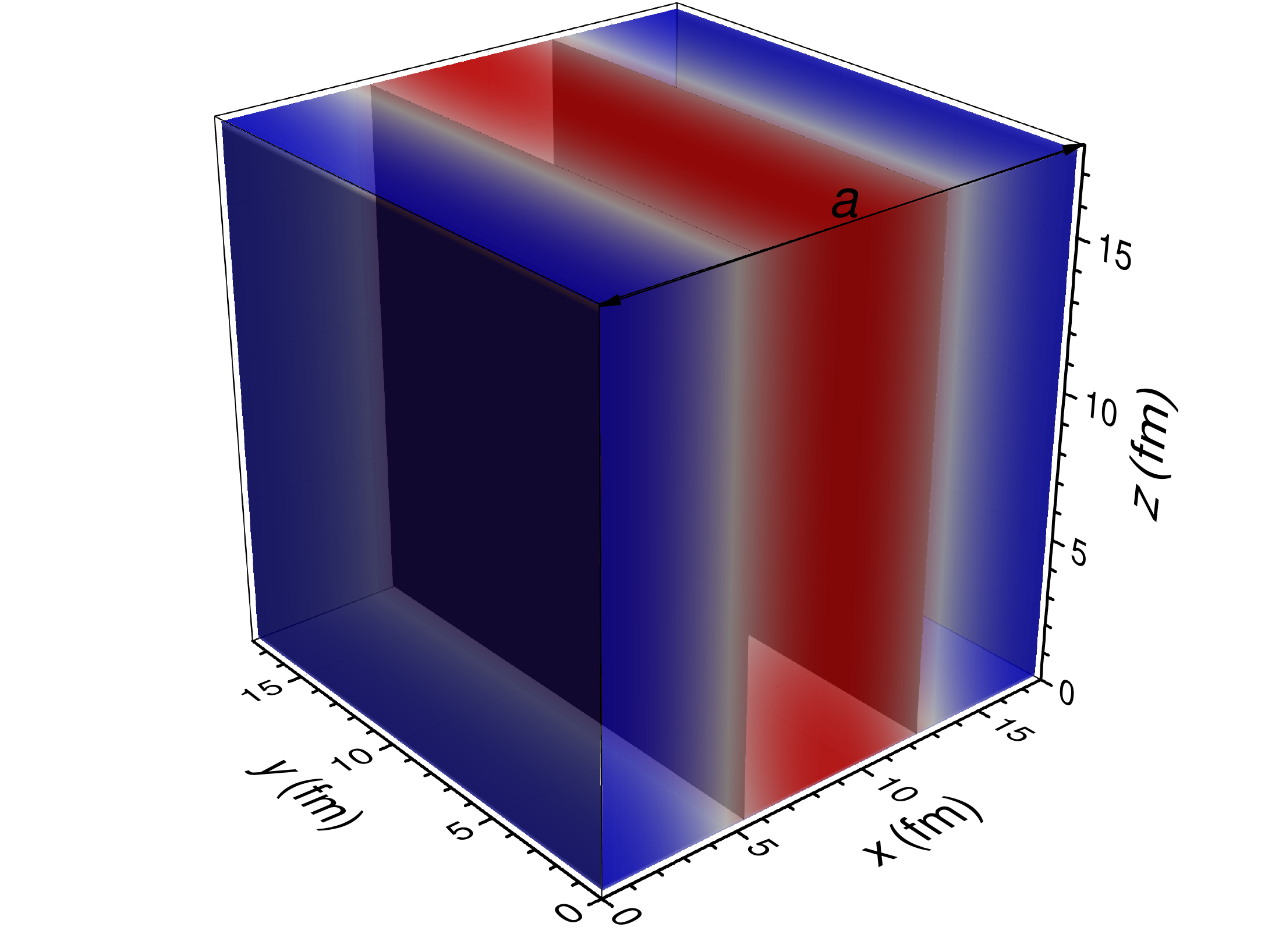}
(e)
\end{minipage}
\caption{\label{Fig:Lattice} The iso-surfaces (at $n_p = 0.05\ \mathrm{fm}^{-3}$) and electron density profiles ($n_e$) of nuclear pasta ($Y_p = 0.5$) in unit cells of typical lattice structures, i.e., (a) Droplets in BCC lattice at $n_\mathrm{b} = 0.01\ \mathrm{fm}^{-3}$; (b) Droplets in FCC lattice at $n_\mathrm{b} = 0.01\ \mathrm{fm}^{-3}$; (c) Rods in honeycomb configuration at $n_\mathrm{b} = 0.03\ \mathrm{fm}^{-3}$, where the lattice constants $b = \sqrt{3}a$; (d) Rods in simple configuration at $n_\mathrm{b} = 0.03\ \mathrm{fm}^{-3}$; (e) Slabs at $n_\mathrm{b} = 0.06\ \mathrm{fm}^{-3}$. The parameter Set 1 listed in Tab.~\ref{tab:NM} is adopted for the isovector channel of effective $N$-$N$ interactions, where the corresponding properties are indicated in Fig.~\ref{Fig:EpA_Set1}. }
\end{figure*}

Adopting the two parameter sets introduced in Sec.~\ref{sec:num_sym}, nuclear pasta structures with the droplets/bubbles forming SC, BCC, and FCC lattices, the rods/tubes forming simple and honeycomb configurations, and slabs are investigated based on Thomas-Fermi approximation, where the numerical details are introduced in Sec.~\ref{sec:num_pasta}. In principle, we should examine all possible pasta structures and search for the optimum one, while only a limited number of nuclear shapes are considered here. Nevertheless, according to previous investigations with random initial density profiles~\cite{Okamoto2012_PLB713-284, Okamoto2013_PRC88-025801}, the pasta structures considered here are likely more stable than other exotic structures.

In Fig.~\ref{Fig:Lattice} we present the typical lattice structures of droplets/rods/slabs obtained in our calculation, while the density profiles of electrons are reversed for bubbles and tubes. For spherical droplets/bubbles forming SC, BCC, and FCC lattices, the corresponding unit cell is a cubic box. As indicated in Fig.~\ref{Fig:Lattice}, the lattice constants on each axis take a same value, i.e., $a=b=c$. If we adopt a cuboid unit cell instead of the cubic one, the BCC lattice can evolve in to a FCC lattice by elongation, i.e., $c=\sqrt{2}a$ and $a=b$ so that the structure (a) takes up half the volume of (b) in Fig.~\ref{Fig:Lattice}. For the rod/tube phases in honeycomb configuration, the lattice constants on $x$- and $y$-axis take different values. By minimizing the energy per baryon with respect to $a$ and $b$, we have found $b = \sqrt{3}a$, which is consistent with typical honeycomb configurations. Note that for rod/tube phases, the lattice constant $c$ on $z$-axis has nothing to do with nuclear pasta structures, and the same for the lattice constants $b$ and $c$ in the slab phase.

For SC lattices, the obtained energy per baryon is typically a few keV larger than that of BCC and FCC lattices, we thus disregard the SC lattices here. Note that in the density regions with stable slab and tube phases, we have found the network-like double P surface~\cite{Schuetrumpf2019_PRC100-045806} by assuming BCC lattice as initial configurations. The corresponding energy per baryon is found to be around 0.1 MeV larger than that of ground states. A through investigation on all possible isomeric structures should be carried out in our future works.

\subsection{\label{sec:pasta_fix} Nuclear pasta with fixed proton fraction}
We first investigate the properties of nuclear pasta at fixed proton number fractions, i.e., symmetric nuclear matter with $Y_p =0.5$, and asymmetric nuclear matter with $Y_p =0.3$ and 0.1. Note that for neutron star matter with $\beta$-equilibrium, as will be illustrated in Sec.~\ref{sec:pasta_beta}, the proton fraction becomes much smaller than 0.1.

\begin{figure*}[htbp]
\begin{minipage}[t]{0.34\linewidth}
\centering
\includegraphics[width=\textwidth]{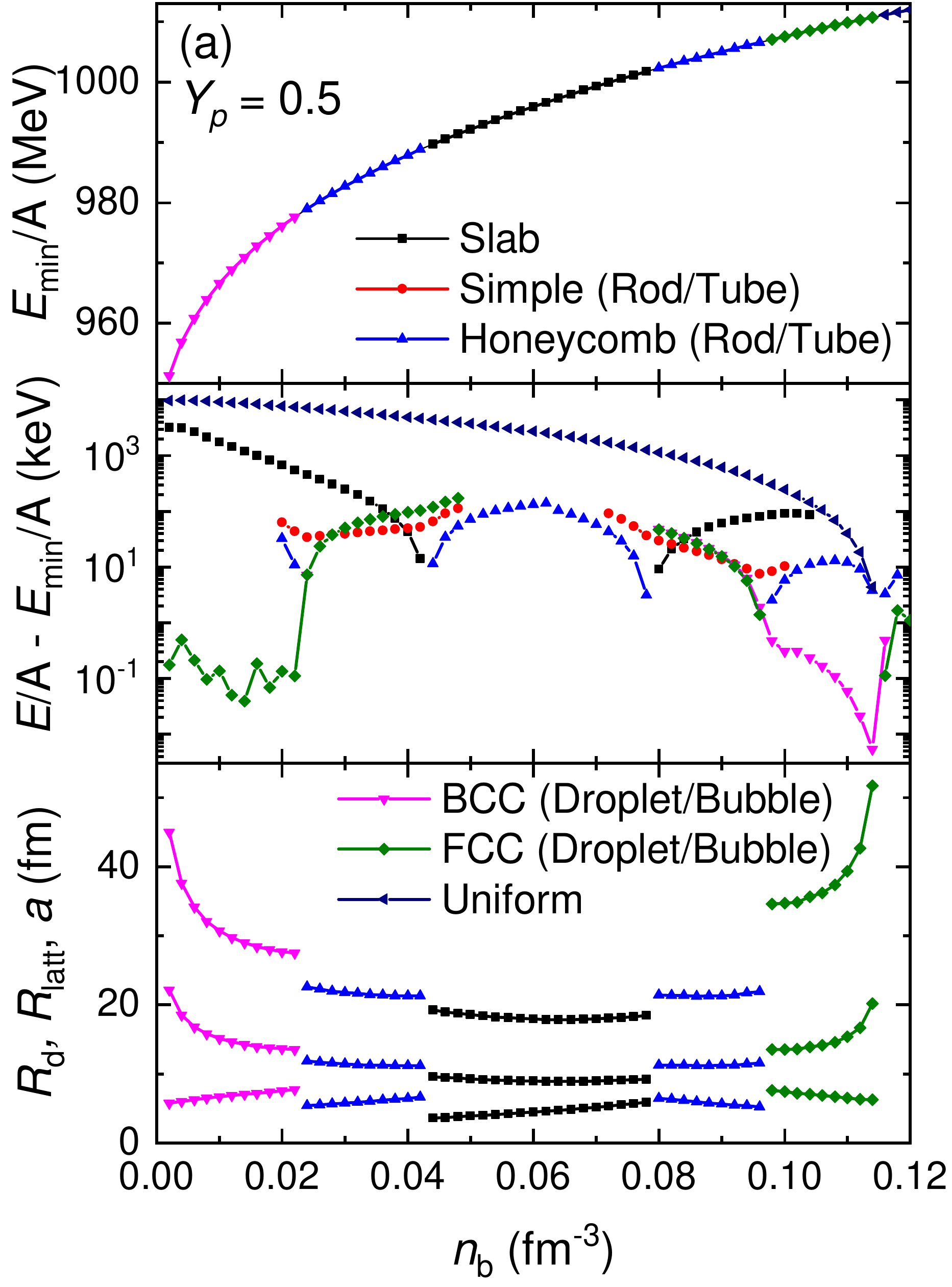}
\end{minipage}%
\hfill
\begin{minipage}[t]{0.322\linewidth}
\centering
\includegraphics[width=\textwidth]{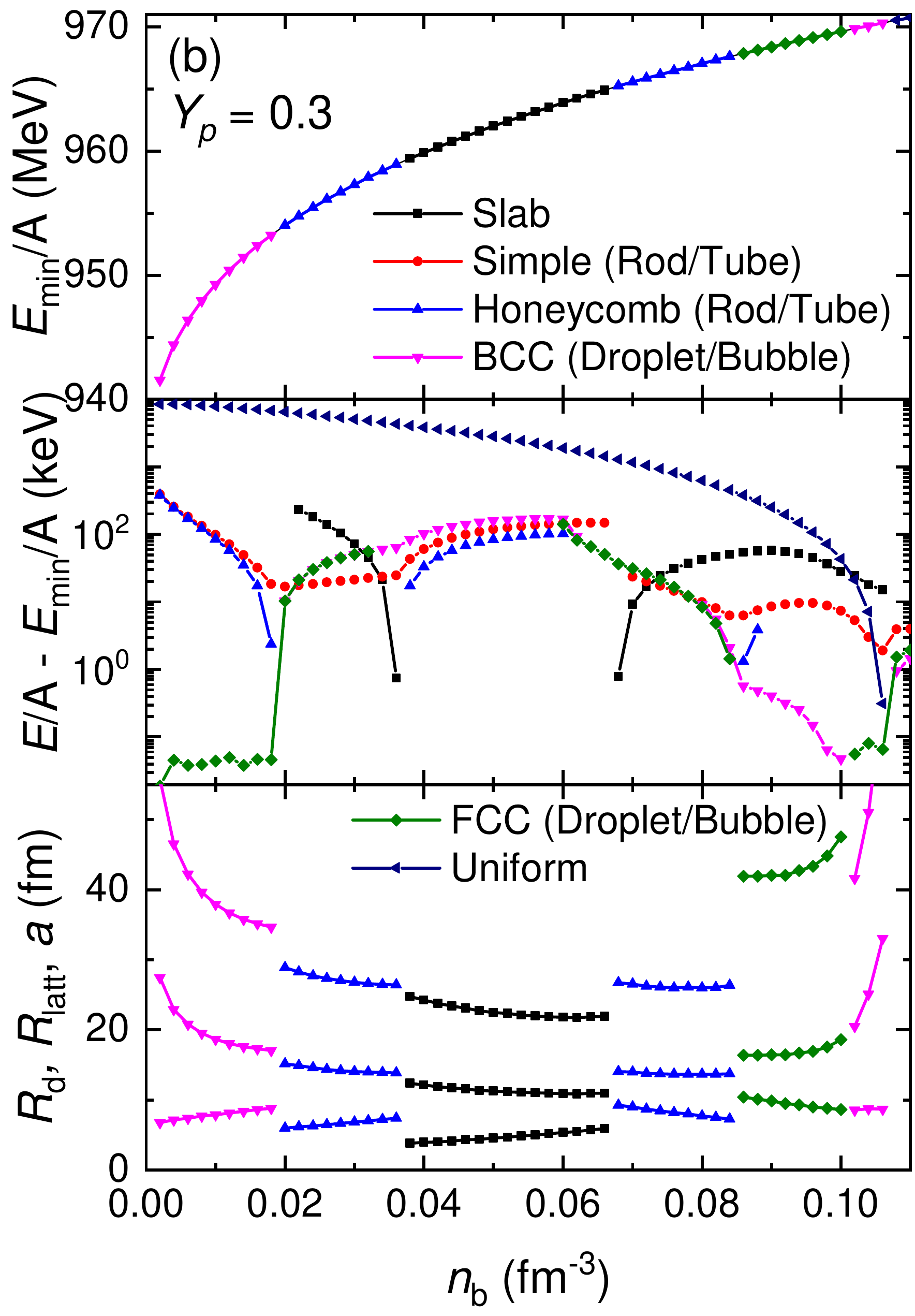}
\end{minipage}
\hfill
\begin{minipage}[t]{0.325\linewidth}
\centering
\includegraphics[width=\textwidth]{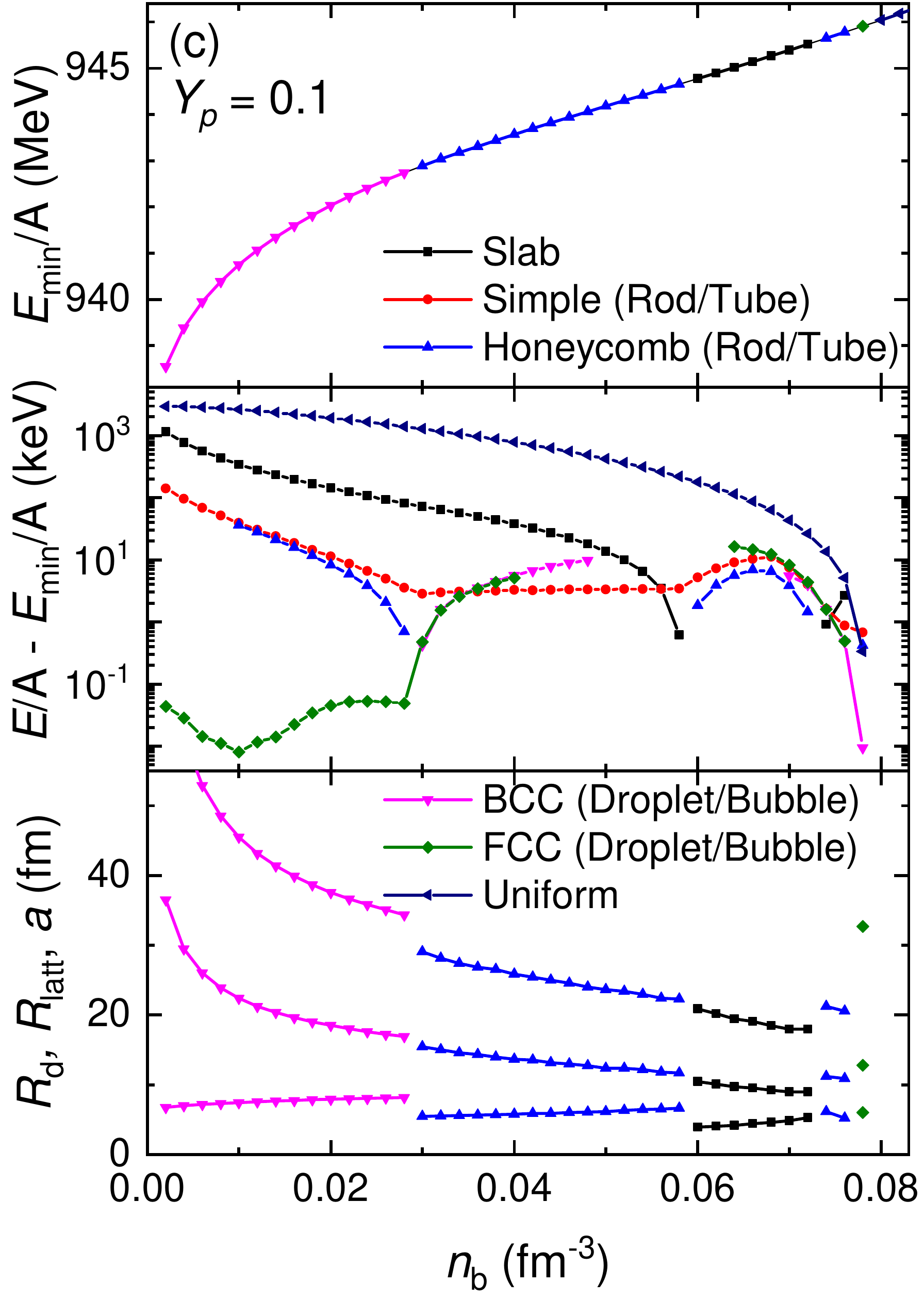}
\end{minipage}
\caption{\label{Fig:EpA_Set0} Minimum energy per baryon, energy excess per baryon, droplet size $R_\mathrm{d}$ and lattice constants ($R_\mathrm{latt}$ and $a$) for nuclear matter with proton fractions $Y_p =0.5$, 0.3, and 0.1. The parameter Set 0 listed in Tab.~\ref{tab:NM} is adopted for the isovector channel of effective $N$-$N$ interactions.}
\end{figure*}

\begin{figure*}[htbp]
\begin{minipage}[t]{0.333\linewidth}
\centering
\includegraphics[width=\textwidth]{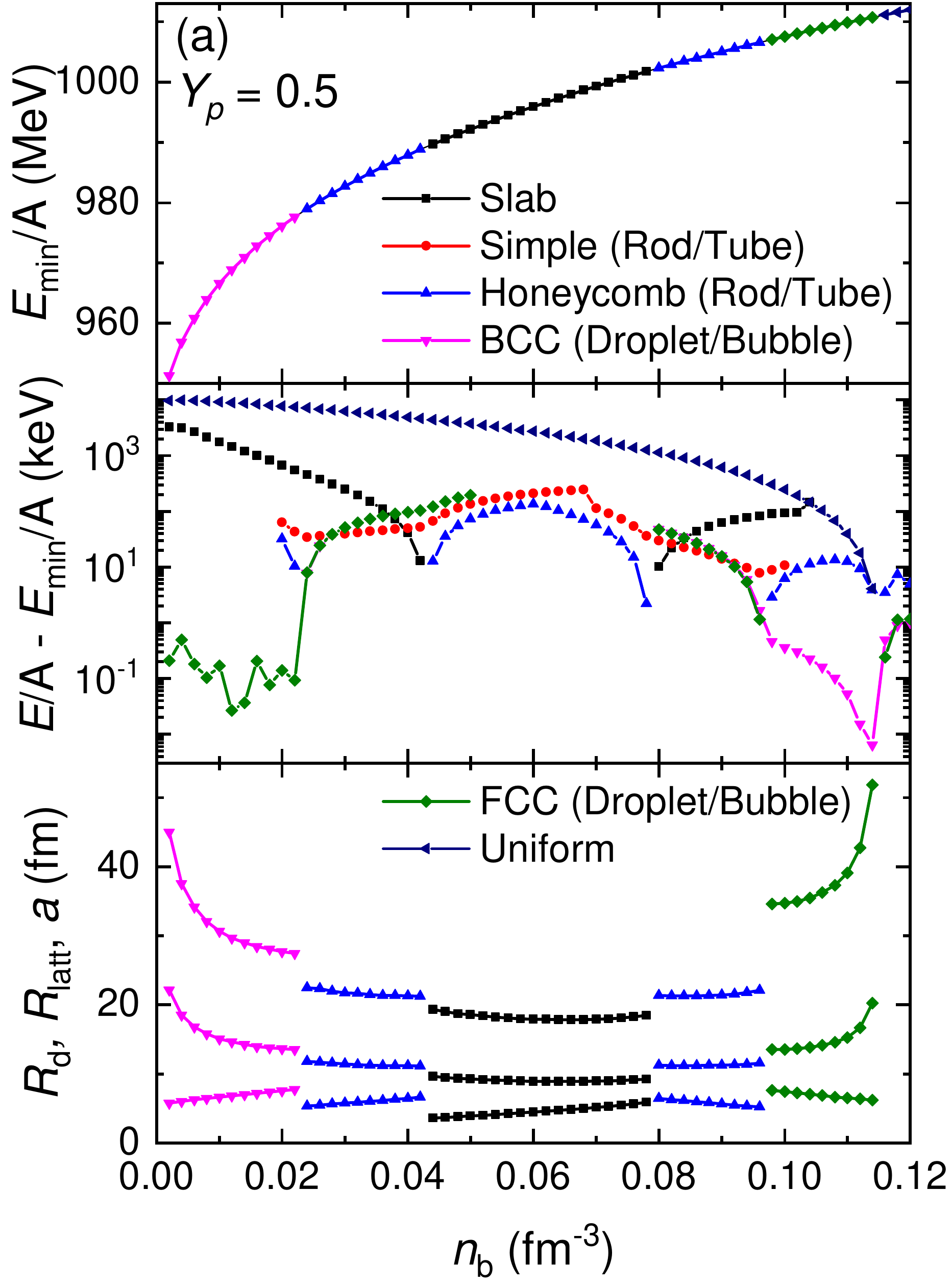}
\end{minipage}%
\hfill
\begin{minipage}[t]{0.321\linewidth}
\centering
\includegraphics[width=\textwidth]{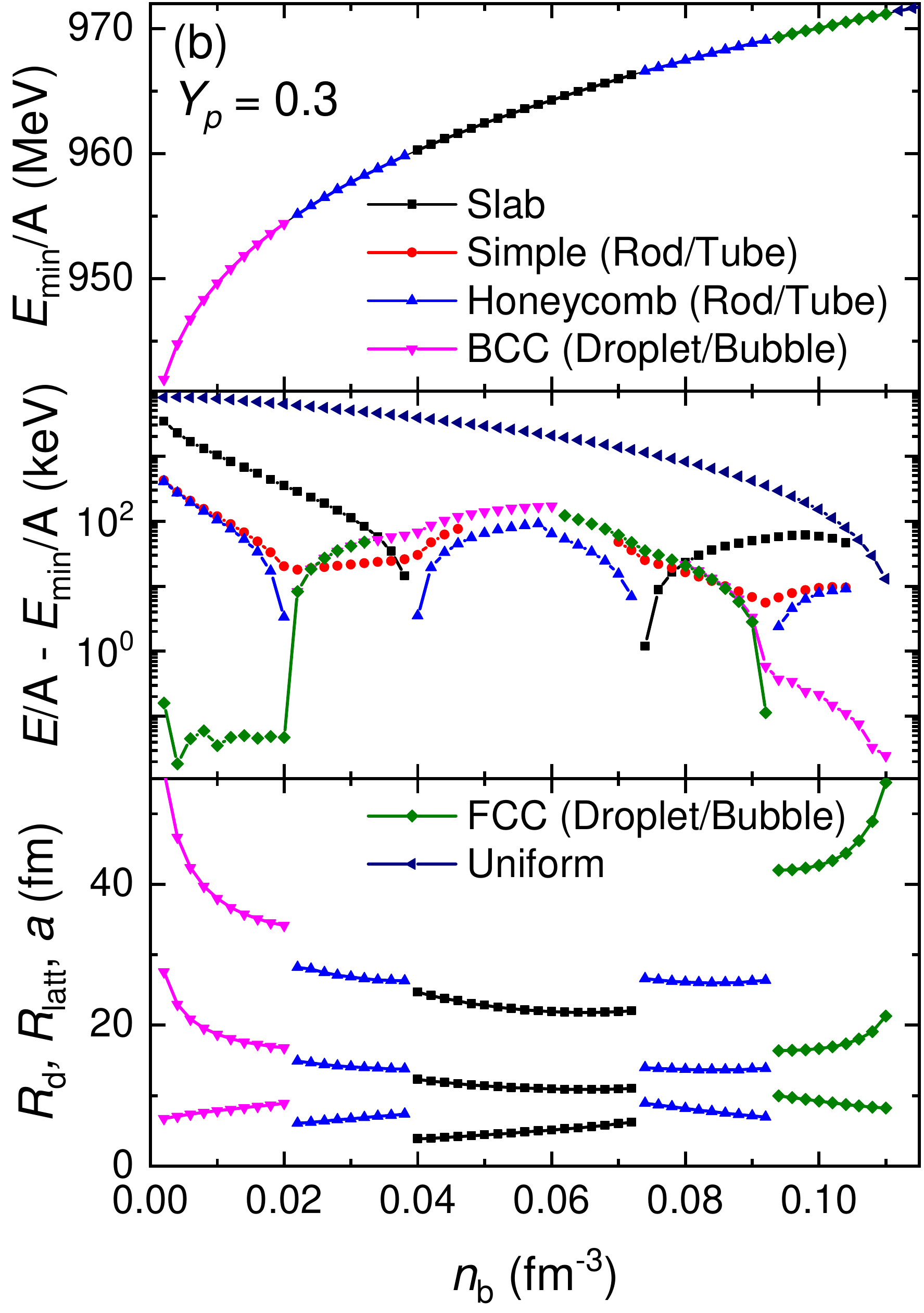}
\end{minipage}
\hfill
\begin{minipage}[t]{0.332\linewidth}
\centering
\includegraphics[width=\textwidth]{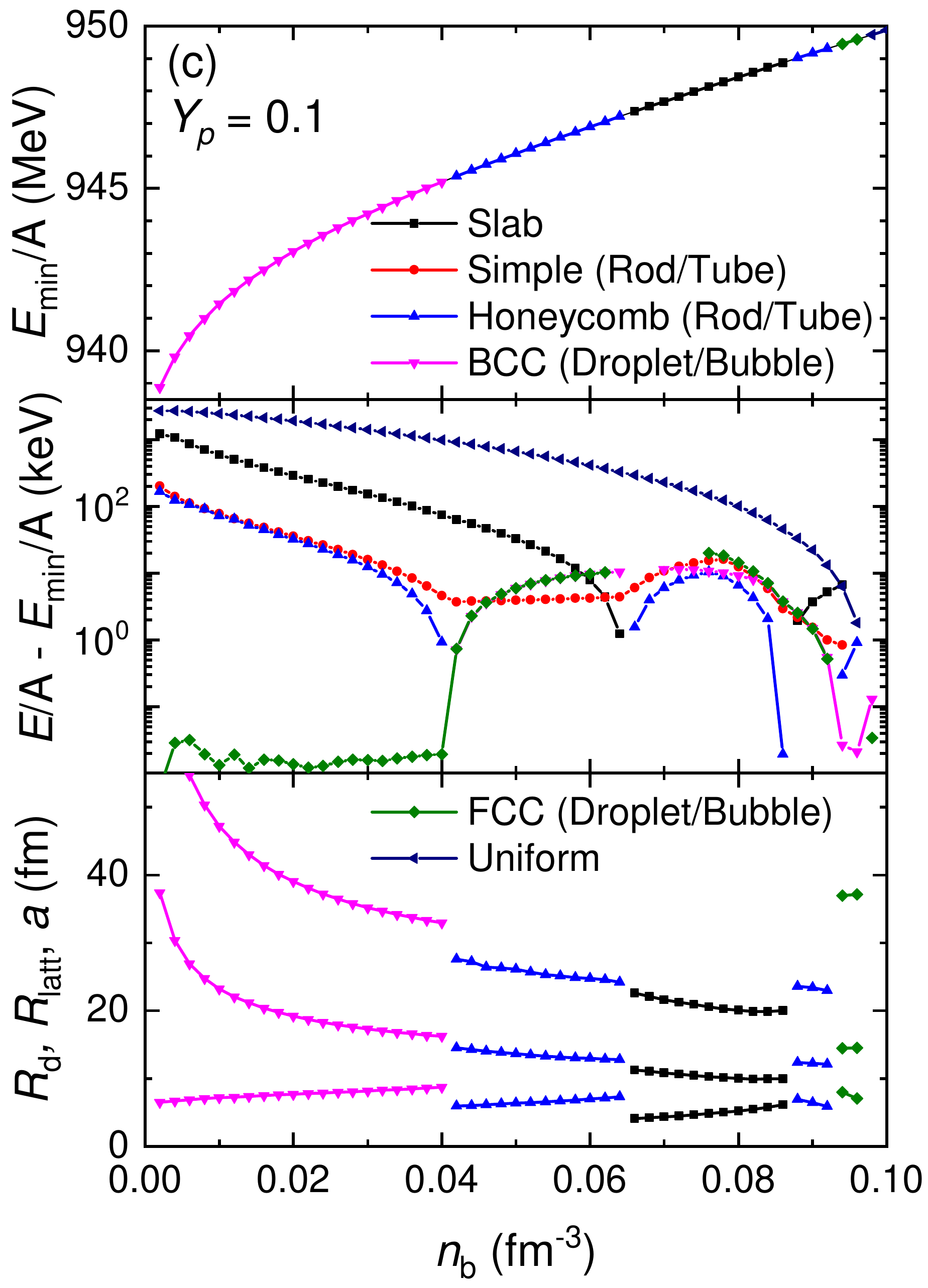}
\end{minipage}
\caption{\label{Fig:EpA_Set1} Same as Fig.~\ref{Fig:EpA_Set0} but adopting the parameter Set 1 in Tab.~\ref{tab:NM}.}
\end{figure*}

\begin{figure}
\includegraphics[width=\linewidth]{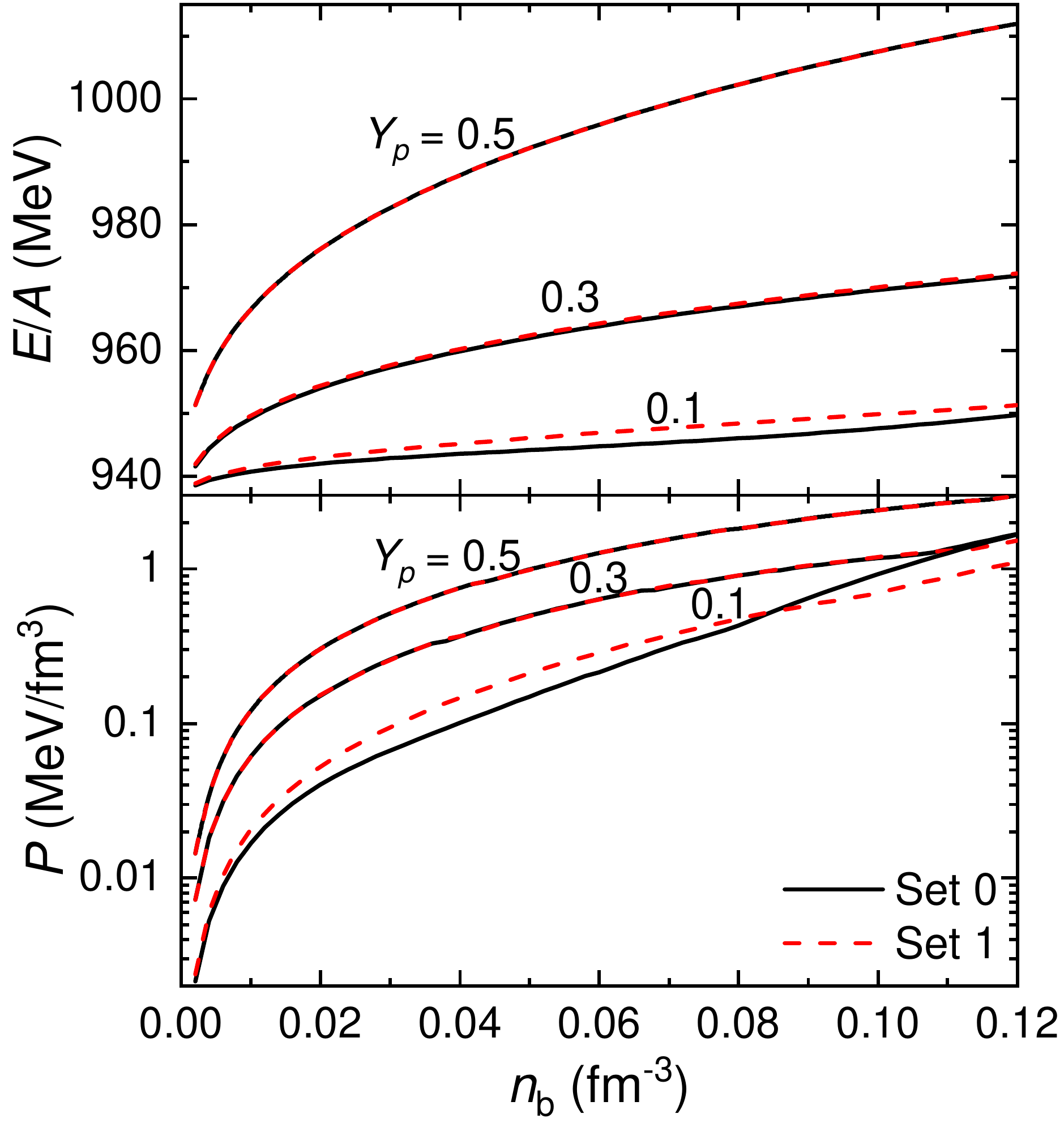}
\caption{\label{Fig:Pressure} Comparison between the energies per baryon and pressures of nuclear matter predicted by the two parameter sets in Tab.~\ref{tab:NM}, corresponding to the most favorable configurations indicated in Figs.~\ref{Fig:EpA_Set0} and~\ref{Fig:EpA_Set1}.}
\end{figure}

In Figs.~\ref{Fig:EpA_Set0} and~\ref{Fig:EpA_Set1}, we present the obtained energy per baryon, droplet size $R_\mathrm{d}$, and lattice constants ($R_\mathrm{latt}$ and $a$) for nuclear matter in various configurations, where the parameter sets listed in Tab.~\ref{tab:NM} are adopted for the isovector channel of effective $N$-$N$ interactions. The energy per baryon and pressure corresponding to the most favorable configurations are indicated in Fig.~\ref{Fig:Pressure}, where the energies per baryon of symmetric nuclear matter ($Y_p =0.5$) are indistinguishable between the values obtained with the two parameter sets. Meanwhile, the parameter Set 1 with $L = 41.34$ MeV predicts larger energy per baryon for asymmetric nuclear matter, which is mainly due to a larger symmetry energy at subsaturation densities. Consequently, only the pressures of asymmetric nuclear matter ($Y_p =0.1$) are altered, where Set 1 predicts softer EOSs, i.e., larger $P$ at $n_\mathrm{b}\lesssim 0.1\ \mathrm{fm}^{-3}$ and smaller $P$ at higher densities due to a smoother behavior of symmetry energy. Similar situation is also expected in $\beta$-stable matter, which reduces the radii, tidal deformability, and maximum mass of neutron stars as indicated in Fig.~\ref{Fig:MRL}.

Comparing with the uniform phase, the energy per baryon decreases by up to 10 MeV with the emergence of nonuniform structures. As density increases, the most favorable configuration changes from the droplets in BCC lattice to rods in honeycomb lattice, slabs, tubes in honeycomb lattice, bubbles in FCC lattice, and to the uniform phase. The energy excess per baryon with respect to different configurations are indicated in the center panels of Figs.~\ref{Fig:EpA_Set0} and~\ref{Fig:EpA_Set1}. In contrast to the SC lattice, it is found that throughout the density region, the obtained energies per baryon are rather close to each other for droplets/bubbles in both BCC and FCC lattices, where the differences lie within $\sim$0.1 keV. Such a small difference is consistent with the analytical estimations in Ref.~\cite{Oyamatsu1984_PTP72-373}. However, this makes it difficult for us to distinguish between the two lattice configurations, especially in the cases with large unit cells. For the droplet phases, we find that the BCC lattice is more stable than FCC lattice. For the bubble phases, on the contrary, FCC lattice is more stable. The energy difference between the bubble phases in two lattice configurations decreases with density and BCC lattice may become more stable, e.g., the BCC lattice appeared in between the FCC lattice and the uniform phase for the asymmetric nuclear matter ($Y_p =0.3$) as indicated in Fig.~\ref{Fig:EpA_Set0}. Meanwhile, we notice that the rod phases in simple lattice are always a few keV larger than that of the honeycomb lattice, while the energy differences between droplets, rods, slabs, tubes, bubbles, and uniform matter are more evident.

\begin{figure}
\includegraphics[width=\linewidth]{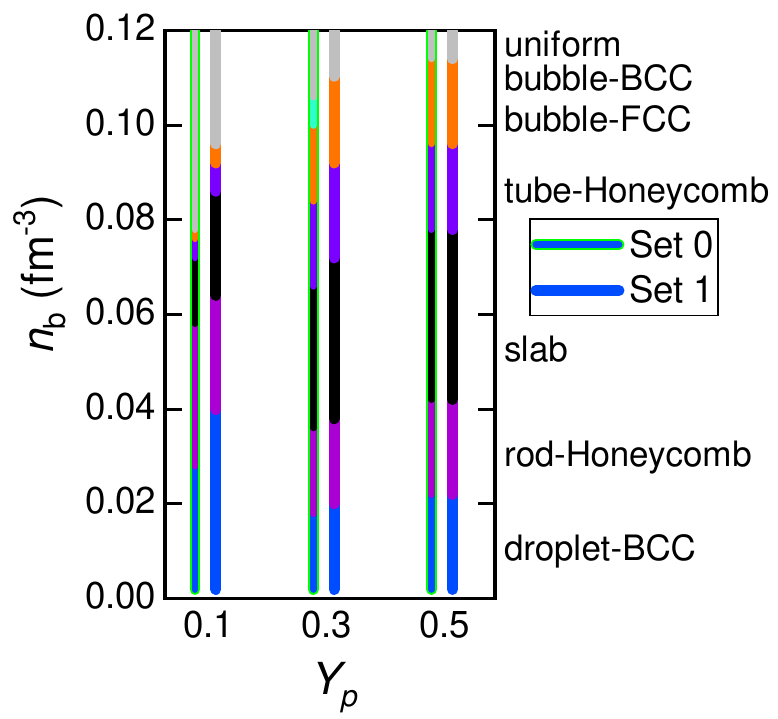}
\caption{\label{Fig:Phase} Phase diagrams of nuclear pasta obtained with parameter Set 0 (left) and Set 1 (right) in Tab.~\ref{tab:NM}.}
\end{figure}

A detailed comparison of the phase diagrams obtained with both parameter sets are presented in Fig.~\ref{Fig:Phase}, which are identical for the cases with symmetric nuclear matter ($Y_p =0.5$). The distinction between different slopes of symmetry energy starts to take place for asymmetric nuclear matter ($Y_p =0.1$ and 0.3), where the core-crust transition density and the onset density of non-spherical nuclei become larger for Set 1 with smaller $L$. This is consistent with previous studies~\cite{Oyamatsu2007_PRC75-015801, Grill2012_PRC85-055808, Bao2015_PRC91-015807, Shen2020_ApJ891-148}. Note that the phase diagrams of Set 0 are slightly different from our previous investigations~\cite{Okamoto2012_PLB713-284, Okamoto2013_PRC88-025801}, where we have now considered the effect of finite cell size and work with optimum cell sizes as illustrated in Fig.~\ref{Fig:Box}.

The droplet size $R_\mathrm{d}$ and lattice constant $R_\mathrm{latt}$ are related to the corresponding sizes in spherical and cylindrical approximations of the WS cell, which are obtained with
\begin{equation}
 \left(R_d\right)^D =
 \left\{\begin{array}{l}
   \left(R_\mathrm{latt}\right)^D \frac{\langle n_p \rangle^2}{\langle n_p^2 \rangle},  \text{\ \ \ \ \ \ \ \ \ \ \ droplet-like}\\
   \left(R_\mathrm{latt}\right)^D \left(1- \frac{\langle n_p \rangle^2}{\langle n_p^2 \rangle}\right),  \text{\ \ bubble-like}\\
 \end{array}\right.,  \label{Eq:Rd}
\end{equation}
and
\begin{equation}
  \left(R_\mathrm{latt}\right)^D =
 \left\{\begin{array}{l}
   \frac{3}{4\pi N_d} {\Delta x N_x \Delta y N_y \Delta z N_z},\  D = 3\\
   \frac{1}{\pi N_d}\Delta x N_x \Delta y N_y, \ \ \ \ \ \ \ \ \   D = 2\\
   \frac{1}{2 N_d}\Delta x N_x, \ \ \ \ \ \ \ \ \ \ \ \ \ \ \ \  D = 1\\
 \end{array}\right.. \label{Eq:Rlatt}
\end{equation}
Here $N_d$ represents the number of droplets in one octant of the unit cell, and $D$ is the dimension with $D = 3$ for droplets and bubbles, $D = 2$ for rods and tubes, and $D = 1$ for slabs. As indicted in the bottom panels of Figs.~\ref{Fig:EpA_Set0} and~\ref{Fig:EpA_Set1}, the obtained lattice constant $R_\mathrm{latt}$ in spherical and cylindrical approximations of the WS cell fulfills the relation $a \approx 2 R_\mathrm{latt}$ for BCC and honeycomb configurations, while for the slab phase the relation is exactly fulfilled. For the FCC lattice, $a$ is much larger than that of BCC lattice. Nevertheless, we find that the droplet sizes $R_\mathrm{d}$ and lattice constants $R_\mathrm{latt}$ are indistinguishable between BCC and FCC configurations. In fact, the optimum volume of the FCC lattice is twice the BCC lattice ($a_\mathrm{FCC} = 2^{1/3}a_\mathrm{BCC}$), then at fixed volume the BCC lattice can evolve into FCC lattice by elongation. Since the number of droplets in a FCC unit cell is exactly twice the number in a BCC unit cell, each droplet takes up the same volume, so that $R_\mathrm{latt}$ is indistinguishable between BCC and FCC configurations. In such cases, the lattice structure has little impact on droplet properties. The differences between the energies per baryon of BCC and FCC configurations are thus mainly caused by the differences in the Coulomb energies, which are small in the first place~\cite{Oyamatsu1984_PTP72-373}. In general, $R_\mathrm{d}$, $R_\mathrm{latt}$, and $a$ decrease in the order of droplet/bubble phase, rod/tube phase, and slab phase, which is similar to previous findings, e.g., those in Ref.~\cite{Maruyama2005_PRC72-015802}. For each configuration, its size becomes larger for asymmetric nuclear matter with smaller proton fraction $Y_p$. Comparing with the results predicted by the two parameter sets in Tab.~\ref{tab:NM}, we find $R_\mathrm{d}$, $R_\mathrm{latt}$, and $a$ become slightly larger if parameter Set 0 with larger $L$ are adopted.

\subsection{\label{sec:pasta_beta} Nuclear pasta in $\beta$-equilibrium}

\begin{figure*}[htbp]
\begin{minipage}[t]{0.5\linewidth}
\centering
\includegraphics[width=0.8\textwidth]{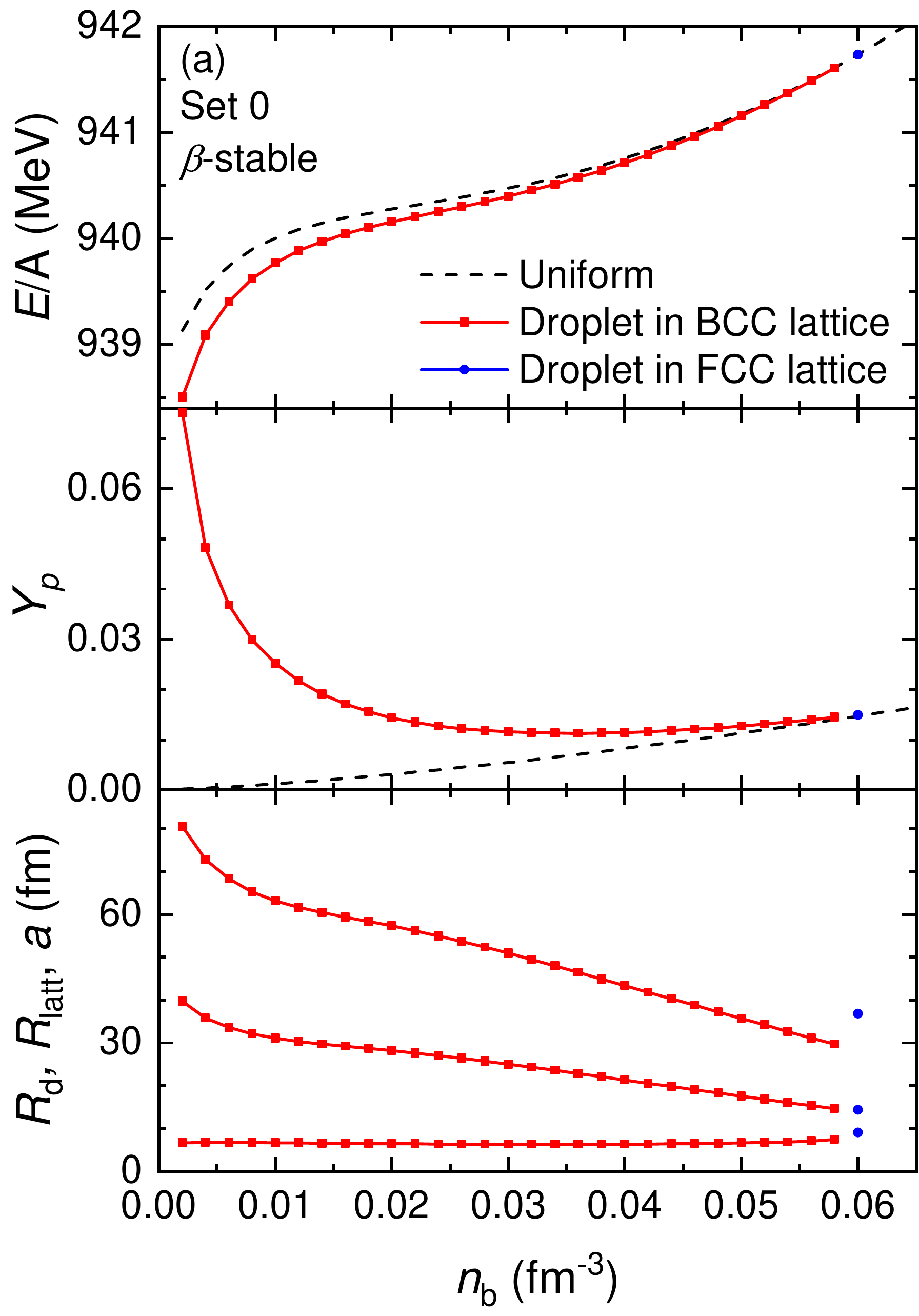}
\end{minipage}%
\hfill
\begin{minipage}[t]{0.5\linewidth}
\centering
\includegraphics[width=0.8\textwidth]{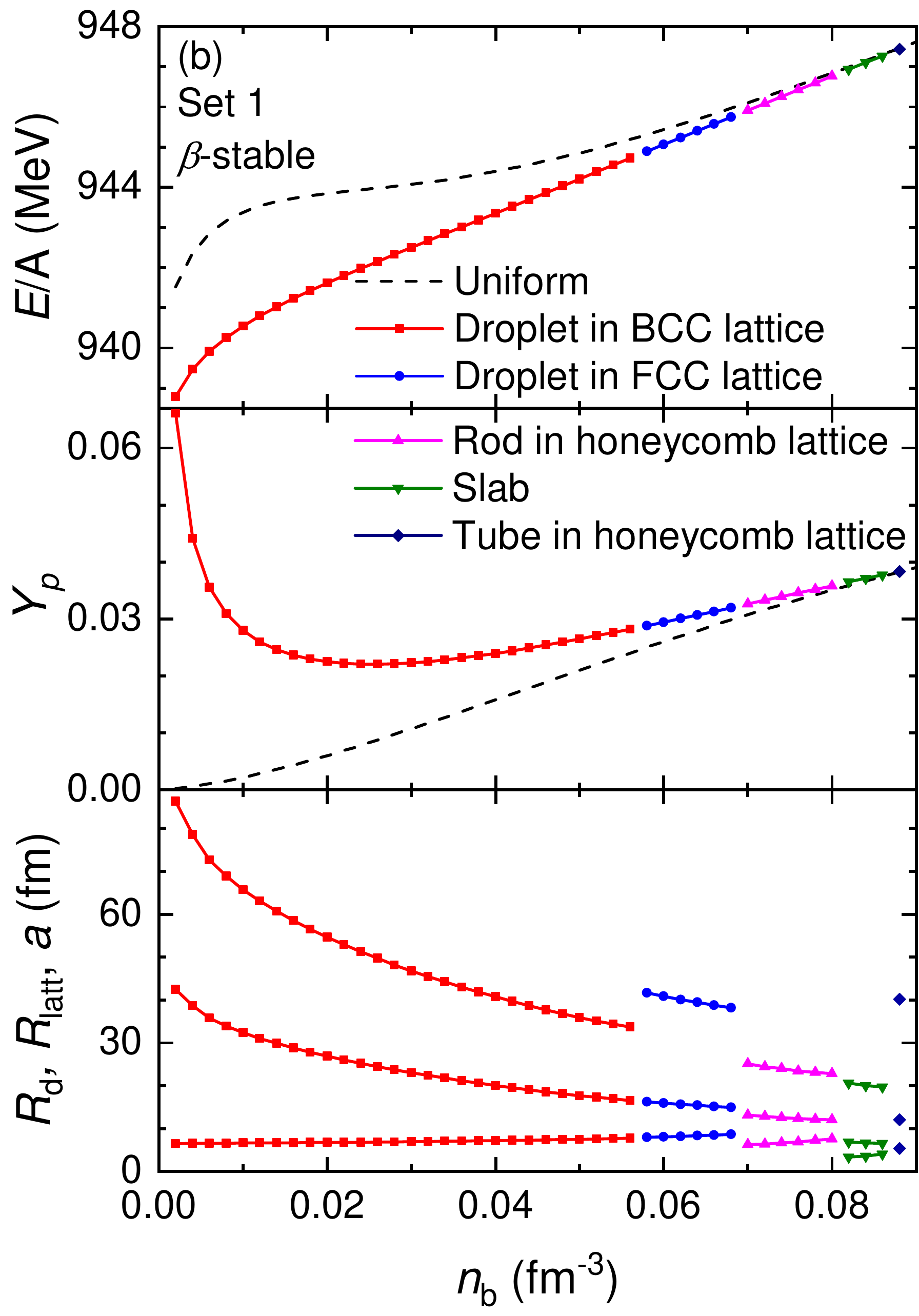}
\end{minipage}
\caption{\label{Fig:EpA_beta} Energy per baryon, proton fraction, droplet size $R_\mathrm{d}$ and lattice constants ($R_\mathrm{latt}$ and $a$) for nuclear matter in $\beta$-equilibrium. The parameter sets 0 (a) and 1 (b) listed in Tab.~\ref{tab:NM} are adopted for the isovector channel of effective $N$-$N$ interactions.}
\end{figure*}

Now we consider the neutron star matter at zero temperature and investigate the nonuniform structures of nuclear matter in $\beta$-equilibrium. In Fig.~\ref{Fig:EpA_beta} we present the energy per baryon, proton fraction, droplet size and lattice constants for the most favorable configurations of nuclear matter in $\beta$-equilibrium. Comparing with the uniform phase, the energy per baryon is reduced by up to 1 MeV with the emergence of nonuniform structures, where the proton fractions increase significantly for nuclear pasta. Meanwhile, we notice that the energy reduction becomes larger if Set 1 is adopted for the isovector channel of effective $N$-$N$ interactions, which corresponds to larger symmetry energies (smaller $L$) at subsaturation densities.

For the phase diagrams of nuclear pasta in $\beta$-equilibrium, only the droplet phases in BCC and FCC lattices emerge if Set 0 is adopted, which is consistent with previous results obtained by adopting spherical approximation for the WS cell~\cite{Maruyama2005_PRC72-015802}. Meanwhile, if Set 1 is adopted, rods/tubes in honeycomb configuration and slabs also appear. The core-crust transition density becomes larger as well. In such cases, a smaller slope of symmetry energy $L$ favors the nonuniform structures for nuclear matter and consequently increases the density region of crusts in neutron stars, which may play important roles in the glitch activities of pulsars~\cite{Link1999_PRL83-3362, Andersson2012_PRL109-241103, Li2016_ApJS223-16, Watanabe2017_PRL119-062701}.

Similar to our findings with fixed proton fractions, the obtained values of $R_\mathrm{d}$, $R_\mathrm{latt}$, and $Y_p$ are indistinguishable between BCC and FCC configurations, while the lattice constant $a$ of FCC lattice is larger than that of BCC lattice. Meanwhile, we notice that $R_\mathrm{d}$, $R_\mathrm{latt}$, and $Y_p$ usually decrease with density, as the proton fraction varies continuously. Finally, the EOS table corresponding to the obtained neutron star matter in $\beta$-equilibrium is presented in Tab.~\ref{table:EOS}. Note that our results for Set 0 are indistinguishable from the previous one in \cite[Table II]{Maruyama2005_PRC72-015802}.

\begin{table}
\caption{\label{table:EOS} The pressure $P$, energy density $E/V$, and proton number ratio $Y_p$ of neutron star matter in $\beta$-equilibrium, where the parameter Set 1 listed in Tab.~\ref{tab:NM} is adopted for the isovector channel of effective $N$-$N$ interactions. For the EOS table obtained with Set 0, please refer to \cite[Tab. II]{Maruyama2005_PRC72-015802}.}
\begin{tabular}{c|cccc} \hline \hline
   & $n_\mathrm{b}$ &$Y_p$ & $E/V$          &     $P$         \\
   & fm${}^{-3}$    &      & MeV/fm${}^{3}$ & MeV/fm${}^{3}$  \\ \hline
\multirow{34}{*}{Droplet}
   & 0.002  &  0.06612 &  1.87757   &    0.00188      \\
   & 0.004  &  0.04412 &  3.75788   &    0.00421      \\
   & 0.006  &  0.03554 &  5.63948   &    0.00690      \\
   & 0.008  &  0.03092 &  7.52206   &    0.00992      \\
   & 0.010  &  0.02798 &  9.40543   &    0.01329      \\
   & 0.012  &  0.02599 &  11.28953  &	0.01705      \\
   & 0.014  &  0.02460 &  13.17428  &    0.02126      \\
   & 0.016  &  0.02363 &  15.05968  &	0.02598      \\
   & 0.018  &  0.02294 &  16.94569  &	0.03126      \\
   & 0.020  &  0.02250 &  18.83233  &	0.03716      \\
   & 0.022  &  0.02222 &  20.71959  &	0.04373      \\
   & 0.024  &  0.02209 &  22.60748  &	0.05099      \\
   & 0.026  &  0.02207 &  24.49601  &	0.05898      \\
   & 0.028  &  0.02216 &  26.38518  &	0.06771      \\
   & 0.030  &  0.02232 &  28.27499  &    0.07721     \\
   & 0.032  &  0.02253 &  30.16547  &	0.08750      \\
   & 0.034  &  0.02281 &  32.05662  &	0.09856      \\
   & 0.036  &  0.02316 &  33.94843  &	0.11038      \\
   & 0.038  &  0.02354 &  35.84093  &	0.12298      \\
   & 0.040  &  0.02396 &  37.73411  &	0.13634      \\
   & 0.042  &  0.02441 &  39.62797  &	0.15046      \\
   & 0.044  &  0.02490 &  41.52253  &	0.16531      \\
   & 0.046  &  0.02542 &  43.41777  &    0.18088     \\
   & 0.048  &  0.02581 &  45.31371  &    0.19729   \\
   & 0.050  &  0.02647 &  47.21035  &	0.21417      \\
   & 0.052  &  0.02692 &  49.10767  &	0.23194      \\
   & 0.054  &  0.02764 &  51.00569  &	0.25010      \\
   & 0.056  &  0.02823 &  52.90440  &    0.26905      \\   \cline{2-5}
   & 0.058 & 0.02883 &	54.80379&	0.28861     \\
   & 0.060 & 0.02944 &	56.70387&	0.30877     \\
   & 0.062 & 0.03006 &	58.60463&	0.32951     \\
   & 0.064 & 0.03068 &	60.50607&	0.35081     \\
   & 0.066 & 0.03130 &  62.40819&	0.37265     \\
   & 0.068 & 0.03193 &	64.31097&	0.39502     \\   \hline
\multirow{6}{*}{Rod}
   & 0.070 & 0.03266 & 66.21441	 & 0.41706    \\
   & 0.072 & 0.03329 & 68.11849  & 0.44050    \\
   & 0.074 & 0.03393 & 70.02324  & 0.46440    \\
   & 0.076 & 0.03456 & 71.92863  & 0.48873    \\
   & 0.078 & 0.03519 & 73.83468  & 0.51353    \\
   & 0.080 & 0.03582 & 75.74136	 & 0.53877     \\  \hline
\multirow{3}{*}{Slab}
   & 0.082 & 0.03653&	77.64866&	0.56238  \\
   & 0.084 & 0.03716&	79.55656&	0.58888     \\
   & 0.086 & 0.03777&	81.46510&   0.61545   \\ \hline
\multirow{1}{*}{Tube}
   & 0.088 &	0.03836&	83.37425&	0.64035    \\   \hline
\multirow{6}{*}{Uniform}
   & 0.090 & 0.03898	&85.28397	&0.66675    \\
   & 0.092 & 0.03971	&87.19433	&0.69753    \\
   & 0.094 & 0.04042	&89.10536	&0.72917    \\
   & 0.096 & 0.04111	&91.01708	&0.76170    \\
   & 0.098 & 0.04179	&92.92948	&0.79512    \\
   & 0.100 & 0.04245	&94.84257	&0.82949    \\
\hline
\end{tabular}
\end{table}

\section{\label{sec:con}Conclusion}
In this work we have investigated nuclear pasta structures in a three-dimensional geometry with reflection symmetry, where the RMF model with Thomas-Fermi approximation was adopted. To improve the numerical accuracy and efficiency, we have exploited the reflection symmetry of unit cells and expanded the mean fields according to fast cosine transformation, where the computation time was reduced by considering only one octant of the unit cell~\cite{Newton2009_PRC79-055801}. For fixed nuclear shape, lattice structure, baryon number density $n_\mathrm{b}$, and proton fraction $Y_p$, the effect of finite cell size~\cite{GimenezMolinelli2014_NPA923-31, Newton2009_PRC79-055801} was treated carefully by searching for the minimum energy per baryon with respect to the cell size. For droplet/bubble phases, it is found that the obtained energy per baryon in SC lattices is typically a few keV larger than that of BCC and FCC lattices. For rod/tube phases, the energy per baryon for those in simple lattice is always a few keV larger than that of the honeycomb lattice. Meanwhile, we have noticed that the energy per baryon $E/A$, droplet size $R_\mathrm{d}$ and lattice constant $R_\mathrm{latt}$ for droplets/bubbles in BCC and FCC lattices are rather close to each other, suggesting that the properties of droplets/bubbles are insensitive to the lattice structures. The corresponding differences for $E/A$ are found to lie within $\sim$0.1 keV, while the FCC lattice can be obtained by the BCC lattice via elongation. Such a small difference may cause the possible coexistence of both BCC and FCC lattices as polycrystalline configurations.

By introducing an $\omega$-$\rho$ cross coupling term, the slope of symmetry energy was reduced from $L = 89.39$ MeV to  $L = 41.34$ MeV, which is consistent with recent constrains from nuclear physics and pulsar observations~\cite{Zhu2018_ApJ862-98, Tsang2019_PLB795-533, Dexheimer2019_JPG46-034002, Zhang2019_EPJA55-39, Tsang2019_PLB795-533, Zhang2020_PRC101-034303, Li2020_PRC102-045807}. More specifically, the neutron skin thickness of $^{208}$Pb, radii and tidal deformation of 1.4 solar-mass neutron stars coincide with recent observations as the slope of symmetry energy is reduced to $L = 41.34$ MeV. The impact of adopting different slopes of symmetry energy is then examined for nuclear pasta structures with both fixed proton fractions and $\beta$-equilibration. For symmetric nuclear matter, as expected, the difference is insignificant. However, for asymmetric nuclear matter, the obtained core-crust transition density and the onset density of non-spherical nuclei become larger for smaller $L$, which confirms previous findings adopting the spherical and cylindrical approximation for the WS cell~\cite{Oyamatsu2007_PRC75-015801, Grill2012_PRC85-055808, Bao2015_PRC91-015807, Shen2020_ApJ891-148}. Instead of BCC lattice, stable droplets/bubbles in FCC lattice emerge as density increases, where the density range becomes larger for smaller $L$ as well. Meanwhile, the differences between droplets/bubbles in BCC and FCC lattices are found to be small despite the large difference on $L$. Bulk properties of neutron stars such as their maximum mass are not so changed by the variation of $L$. However, the larger core-crust transition density predicted by smaller $L$ is expected to alter the fraction of crust in neutron stars, which may play important roles in explaining the glitch activities of pulsars~\cite{Link1999_PRL83-3362, Andersson2012_PRL109-241103, Li2016_ApJS223-16, Watanabe2017_PRL119-062701}, the quasi-periodic oscillation frequencies in giant flares of magnetars~\cite{Sotani2012_PRL108-201101, Sotani2016_MNRAS464-3101}, the gravitational waves emitted by millisecond pulsars~\cite{Abbott2020_ApJ902-L21}, etc.

\section*{ACKNOWLEDGMENTS}
C.-J. X. would like to thank Prof. Bao-An Li for fruitful discussions. This work was supported by National Natural Science Foundation of China (Grants No.~11705163, No.~11875052, No.~11525524, and No.~11875323), JSPS KAKENHI (Grants No.~20K03951 and No.~20H04742), National SKA Program of China No.~2020SKA0120300, and Ningbo Natural Science Foundation (Grant No.~2019A610066). The support provided by China Scholarship Council during a visit of C.-J. X. to JAEA is acknowledged.


\newpage

%

\end{document}